\documentclass{emulateapj}

\newcommand{\Spitzer}{{\sl Spitzer}}

\newcommand{\HST}{{\sl HST}}

\newcommand{\Msun}{\mbox{$M_{\sun}$}}

\newcommand{\Lsun}{\mbox{$L_{\sun}$}}
\newcommand{\Rsun}{\mbox{$R_{\sun}$}}

\newcommand{\perpix}{\mbox{pixel$^{-1}$}}

\newcommand{\etal}{et al.}

\newcommand{\kms}{\hbox{km~s$^{-1}$}}

\newcommand{\Kp}{\mbox{$K^{\prime}$}}
\newcommand{\Ks}{\mbox{$K_S$}}

\newcommand{\degs}{\mbox{$^{\circ}$}}
\newcommand{\Mtot}{\mbox{$M_{\rm tot}$}}
\newcommand{\Lbol}{\mbox{$L_{\rm bol}$}}

\newcommand{\Teff}{\mbox{$T_{\rm eff}$}}
\newcommand{\logg}{\mbox{$\log(g)$}}

\newcommand{\Lp}{\mbox{${L^\prime}$}}

\newcommand{\twomassbin}{\hbox{2MASS~J1534$-$2952AB}}
\newcommand{\hdbin}{\hbox{HD~130948BC}}

\newcommand{\lhsint}{\hbox{LHS~2397a}}
\newcommand{\lhsbin}{\hbox{LHS~2397aAB}}
\newcommand{\lhsA}{\hbox{LHS~2397aA}}

\newcommand{\nameintlong}{\hbox{2MASS~J22062280$-$2047058}}

\newcommand{\name}{\hbox{2MASS~J2206$-$2047AB}}
\newcommand{\nameA}{\hbox{2MASS~J2206$-$2047A}}
\newcommand{\nameB}{\hbox{2MASS~J2206$-$2047B}}
\newcommand{\nameint}{\hbox{2MASS~J2206$-$2047}}
\newcommand{\orbit}{\hbox{\tt ORBIT}}

\slugcomment{Submitted to ApJ 2009 May 26; accepted 2009 September 24}

\shorttitle{Dynamical Mass of 2MASS~J2206$-$2047}
\shortauthors{Dupuy \etal}

\begin{document}

\title{Dynamical Mass of the M8+M8 Binary
  2MASS~J22062280$-$2047058AB\altaffilmark{*,\dag,\ddag,**}}

\author{Trent J.\ Dupuy, Michael C.\ Liu,\altaffilmark{1} and Brendan
  P. Bowler}

\affil{Institute for Astronomy, University of Hawai`i, 2680 Woodlawn
  Drive, Honolulu, HI 96822}

\altaffiltext{*}{Some of the data presented herein were obtained at
  the W.M. Keck Observatory, which is operated as a scientific
  partnership among the California Institute of Technology, the
  University of California, and the National Aeronautics and Space
  Administration. The Observatory was made possible by the generous
  financial support of the W.M. Keck Foundation.}

\altaffiltext{\dag}{Based partly on observations made with the
  NASA/ESA {\sl Hubble Space Telescope}, obtained from the data
  archive at the Space Telescope Institute. STScI is operated by the
  association of Universities for Research in Astronomy, Inc. under
  the NASA contract NAS 5-26555.}

\altaffiltext{\ddag}{Based partly on observations obtained under
  program ID GN-2001B-C-2 at the Gemini Observatory, which is operated
  by the Association of Universities for Research in Astronomy, Inc.,
  under a cooperative agreement with the NSF on behalf of the Gemini
  partnership: the National Science Foundation (United States), the
  Science and Technology Facilities Council (United Kingdom), the
  National Research Council (Canada), CONICYT (Chile), the Australian
  Research Council (Australia), Minist\'{e}rio da Ci\^{e}ncia e
  Tecnologia (Brazil) and SECYT (Argentina).}
      
\altaffiltext{\S}{Based partly on observations made with ESO
  Telescopes at the Paranal Observatory under program IDs 071.C-00327
  and 077.C-0062.}

\altaffiltext{1}{Alfred P. Sloan Research Fellow}

\begin{abstract}

  We present Keck laser guide star adaptive optics imaging of the
  M8+M8 binary \name. Together with archival \HST, Gemini-North, and
  VLT data, our observations span 8.3~years of the binary's
  35$^{+6}_{-5}$~year orbital period, and we determine a total
  dynamical mass of 0.15$^{+0.05}_{-0.03}$~\Msun, with the uncertainty
  dominated by the parallax error. Using the measured total mass and
  individual luminosities, the Tucson and Lyon evolutionary models
  both give an age for the system of 0.4$^{+9.6}_{-0.2}$~Gyr, which is
  consistent with its thin disk space motion derived from the
  Besan\c{c}on Galactic structure model.  Our mass measurement
  combined with the Tucson (Lyon) evolutionary models also yields
  precise effective temperatures, giving 2660$^{+90}_{-100}$~K and
  2640$^{+90}_{-100}$~K (2550$^{+90}_{-100}$~K and
  2530$^{+90}_{-100}$~K) for components A and B, respectively.  These
  temperatures are in good agreement with estimates for other M8
  dwarfs (from the infrared flux method and the M8 mass benchmark
  \lhsA), but atmospheric model fitting of the integrated-light
  spectrum gives hotter temperatures of 2800$\pm$100~K for both
  components.  This modest discrepancy can be explained by systematic
  errors in the atmospheric models or by a slight underestimate of the
  distance (and thus mass and age) of the system.  We also find the
  observed near-infrared colors and magnitudes do not agree with those
  predicted by the Lyon Dusty models, given the known mass of the
  system.

\end{abstract}



\section{Introduction}

Direct mass measurements are a key underpinning of stellar astronomy,
as the characteristics of stars depend more strongly on mass than any
other property. However, there are only a handful of mass measurements
available for the lowest mass stars ($M$~$<$~0.1~\Msun), which largely
comes from work conducted more than a decade ago
\citep[e.g.,][]{1993AJ....106..773H, 2000A&A...364..665S}.  This is
because such measurements were limited by the scarce number of
low-mass objects known, until wide-field optical and infrared surveys
enabled the discovery of hundreds more
\citep[e.g.,][]{2000AJ....120.1085G} and high-resolution imaging
campaigns identified dozens of visual binaries among these objects
\citep[e.g.,][]{2002ApJ...567L..53C, 2003AJ....126.1526B}.  Today,
many of these binaries are finally yielding dynamical mass
measurements after years of patient orbital monitoring
\citep[e.g.,][]{2008A&A...484..429S, me-2397a}.

The M8 dwarf \nameintlong\ (hereinafter \nameint) was discovered in
the Two Micron All Sky Survey (2MASS) by \citet{2000AJ....120.1085G}
and was revealed to be a binary by \citet{2002ApJ...567L..53C}.
\citet{2006AJ....132.1234C} measured a trigonometric parallax for the
system of 37.5$\pm$3.4~mas, corresponding to a distance of
26.7$^{+2.6}_{-2.1}$~pc.  We present here a dynamical mass for \name\
based on Keck laser guide star adaptive optics (LGS AO) imaging from
our ongoing orbital monitoring program targeting ultracool
binaries. Combining our Keck data with archival {\sl Hubble Space
  Telescope} (\HST), Very Large Telescope (VLT), and Gemini-North
Telescope images, we measure a total mass of
0.15$^{+0.05}_{-0.03}$~\Msun, with the dominant source of uncertainty
being the 9.1\% error in the parallax, which translates into an
asymmetric $^{+32}_{-22}$\% error in the mass. Despite the relatively
large uncertainty in the mass, our measurement reveals significant
discrepancies between the predictions of evolutionary and atmospheric
models and the observed properties of \name.


\section{Observations \label{sec:obs}}

\subsection{Keck/NIRC2 LGS \label{sec:keck}}

We monitored \name\ using the LGS AO system at the Keck~II Telescope
on Mauna Kea, Hawaii \citep{2006PASP..118..297W, 2006PASP..118..310V},
using the facility near-infrared camera NIRC2 in its narrow
field-of-view mode. At each epoch, we obtained data in one or more
filters covering the standard atmospheric windows from the Mauna Kea
Observatories (MKO) filter consortium \citep{mkofilters1,
  mkofilters2}. The LGS provided the wavefront reference source for AO
correction, with the exception of tip-tilt motion. The LGS brightness,
as measured by the flux incident on the AO wavefront sensor, was
equivalent to a $V$~$\approx$~10.1--10.7~mag star. The tip-tilt
correction and quasi-static changes in the image of the LGS as seen by
the wavefront sensor were measured contemporaneously by a second,
lower bandwidth wavefront sensor monitoring \nameint, which saw the
equivalent of an $R$~$\approx$~15.8--16.2~mag star.

\begin{figure}
\plotone{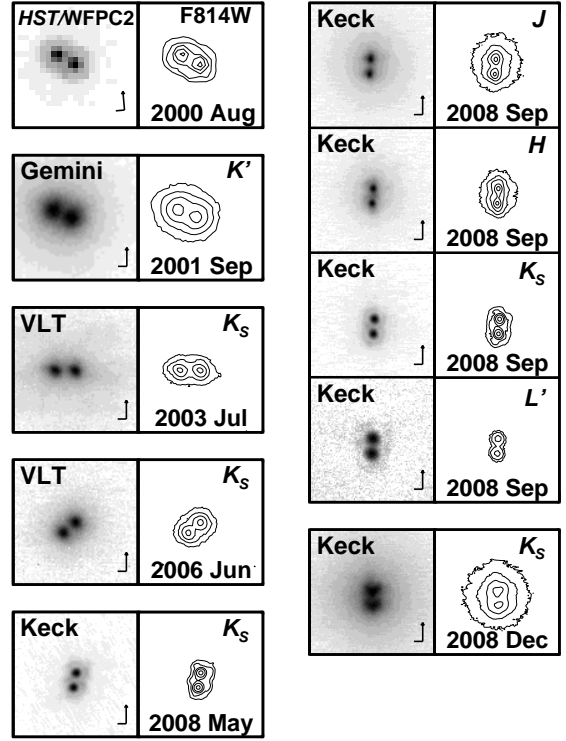}
\caption{ \normalsize \HST, Gemini, VLT, and Keck images of \name\
  shown chronologically by column.  All images are shown on the same
  scale, 1$\farcs$0 on a side, using a square-root stretch for the
  grayscale images.  We do not rotate the \HST\ data so that north is
  up in order to preserve the somewhat undersampled nature of the
  WFPC2 data.  The Airy ring of the Keck PSF is visible in some Keck
  images.  Contours are drawn at 0.78, 0.37, 0.18, 0.085, and 0.040 of
  the peak pixel.  The lowest contour is not drawn for the Gemini,
  VLT, and 2008 December Keck images.  The lowest two contours are not
  drawn for the Keck \Lp\ band images. \label{fig:data}}
\end{figure}

Our procedure for obtaining, reducing, and analyzing our images is
described in detail by \citet{me-2397a}. Table~\ref{tbl:obs}
summarizes our observations of \name, and typical images from each
data set are shown in Figure~\ref{fig:data}. The binary separation,
position angle (P.A.), and flux ratio were determined using the same
three-component Gaussian representation of the point-spread function
(PSF) as described in \citet{me-2397a}. We used the astrometric
calibration from \citet{2008ApJ...689.1044G}, with a pixel scale of
9.963$\pm$0.005~mas~\perpix\ and an orientation for the detector's
$+y$-axis of $+$0$\fdg$13$\pm$0$\fdg$02 east of north, and applied the
distortion correction developed by B. Cameron (2007, private
communication), which changed the results below the 1$\sigma$ level.

To assess systematic errors in our PSF-fitting procedure, we also
applied it to simulated Keck images of \name\ that were created using
images of PSF reference sources with similar FWHM and Strehl,
summarized in Table~\ref{tbl:psf}. After being reduced in an identical
fashion to the science images, the individual dithered images were
shifted with interpolation and added to themselves to match the
observed binary configuration at each epoch.  These simulated images
were then fit in an identical manner to the science images, and the
rms and mean of the truth-minus-fitted parameters determined the
uncertainty and systematic offset.

\tabletypesize{\scriptsize}
\begin{deluxetable}{lccccc}
\tablecaption{Keck LGS AO Observations \label{tbl:obs}}
\tablewidth{0pt}
\tablehead{
\colhead{Date} &
\colhead{Time} &
\colhead{Airmass} &
\colhead{Filter} &
\colhead{FWHM} &
\colhead{Strehl ratio} \\
\colhead{(UT)} &
\colhead{(UT)} &
\colhead{} &
\colhead{} &
\colhead{(mas)} &
\colhead{}}
\startdata
 2008 May 29 & 14:55 & 1.375 &  \Ks\  &  54.7$\pm$0.7  &  0.387$\pm$0.017 \\
 2008 Sep  8 & 08:32 & 1.345 &  $J$   &    52$\pm$3    &  0.050$\pm$0.008 \\
             & 08:27 & 1.352 &  $H$   &    51$\pm$3    &  0.145$\pm$0.015 \\
             & 08:21 & 1.362 &  \Ks\  &  53.6$\pm$1.1  &  0.368$\pm$0.019 \\
             & 08:47 & 1.328 &  \Lp\  &    84$\pm$2    &   0.67$\pm$0.18  \\
 2008 Dec  1 & 05:29 & 1.482 &  \Ks\  &   101$\pm$6\phn&  0.047$\pm$0.010 \\
\enddata
\end{deluxetable}

For the 2008 May and December \Ks\ band data, the simulations
predicted insignificant systematic offsets ($\lesssim$~0.3$\sigma$)
and rms errors that were somewhat smaller than the rms of individual
science dithers. For these data sets, we adopted the science rms
values for the errors in all binary parameters and did not apply the
Monte Carlo offsets. For the 2008 September data ($J$, $H$, \Ks, and
\Lp\ bands), our simulations yielded significant systematic offsets
(as large as 2.4$\sigma$ in \Lp\ band) and errors that were consistent
with or somewhat smaller than the rms of individual science dithers.
Such offsets are expected, particularly in high-Strehl images with
prominent Airy rings, as our multi-Gaussian PSF model is known to be
an imperfect representation of the data.  Because we had data in four
bandpasses at a single epoch, we were able to check that the Monte
Carlo offsets brought the astrometry into better agreement -- if they
did not, our simulations would not have correctly assessed our
systematic errors. We used the Monte Carlo-derived errors to compute
$\chi^2$ for the separation and P.A.  measurements and found that
after applying the systematic offsets $\chi^2$ improved from 17.3 to
2.4 for the separation and from 7.8 to 1.3 for the P.A. (n.b., since
there are 3 degrees of freedom, the median value of $\chi^2$ is
2.4). Thus, we applied the systematic offsets from our Monte Carlo
simulations as they brought the multiband astrometry into agreement,
and we adopted the Monte Carlo errors if they were larger than the rms
of measurements from individual images.

A summary of the astrometry and flux ratios derived from the Keck data
is given in Table~\ref{tbl:astrom}. We used the data set with the
smallest astrometric errors at each epoch in the orbit fit.  As
discussed in Section~\ref{sec:orbit}, the resulting orbit fit changes
insignificantly if we vary the input astrometry by: (1) using the
errors and offsets derived from Monte Carlo simulations rather than
simply using the rms of individual dithers for the errors; (2) using a
different data set for a given epoch.

\tabletypesize{\scriptsize}
\begin{deluxetable}{lccccc}
\tablecaption{Keck PSF Reference Observations \label{tbl:psf}}
\tablewidth{0pt}
\tablehead{
\colhead{Date} &
\colhead{Time} &
\colhead{Airmass} &
\colhead{Filter} &
\colhead{FWHM} &
\colhead{Strehl ratio} \\
\colhead{(UT)} &
\colhead{(UT)} &
\colhead{} &
\colhead{} &
\colhead{(mas)} &
\colhead{}}
\startdata
 2008 May 29\tablenotemark{a} & 13:17 & 1.162 &  \Ks\  &  55.5$\pm$0.8  &  0.395$\pm$0.020 \\
 2008 Sep  8\tablenotemark{b} & 09:15 & 1.333 &  $J$   &    52$\pm$3    &  0.054$\pm$0.003 \\
             & 09:06 & 1.343 &  $H$   &    51$\pm$2    &  0.145$\pm$0.016 \\
             & 09:00 & 1.350 &  \Ks\  &  53.0$\pm$0.9  &  0.348$\pm$0.007 \\
             & 09:21 & 1.328 &  \Lp\  &    84$\pm$2    &   0.42$\pm$0.16  \\
 2008 Dec  1\tablenotemark{c} & 06:16 & 1.329 &  \Ks\  &\phn97$\pm$12   &  0.049$\pm$0.014 \\
\enddata

\tablenotetext{a}{2MASS~J17502484$-$0016151}

\tablenotetext{b}{2MASS~J22345725$-$2101071}

\tablenotetext{c}{2MASS~J21402966$+$1625212}

\end{deluxetable}

\subsection{\HST/WFPC2-PC1  \label{sec:hst}}

\citet{2003AJ....126.1526B} reported binary parameters for \name\
based on their \HST\ discovery images; however, we have chosen to
re-analyze this data because in our previous work we have found that
our PSF fitting technique can yield somewhat more precise astrometry
\citep{2008ApJ...689..436L}. Also, \citet{2003AJ....126.1526B} did not
derive individual measurement errors for each binary in their sample,
and accurate uncertainties are critical for orbit fitting. We
retrieved the \HST\ archival images of \name\ obtained with the WFPC2
Planetary Camera (PC1) on UT 2000 August 13 (GO-8581, PI Reid). These
comprise two 30~s exposures in the F814W bandpass and one 500~s F1042M
exposure. We only used the F814W images for deriving astrometry
because (1) the smaller PSF enables better deblending of this tight
binary and (2) a pair of images offers better cosmic ray rejection
than a single image. We used TinyTim \citep{1995ASPC...77..349K} to
generate PSF models which were fit to the data in a similar fashion to
our previous work \citep{2008ApJ...689..436L, 2009ApJ...692..729D}.
From our PSF fitting of the two images we determined the binary
separation, position angle, and flux ratio.\footnote[2]{We used a pixel
  scale of 45.54$\pm$0.01~mas~\perpix, the quoted pixel scale from the
  WFPC2 Instrument Handbook for Cycle 13, which is consistent with
  other values in the literature \citep[e.g., see the discussion
  by][]{2008ApJ...689..436L}.}

To determine uncertainties and potential systematic offsets for our
measurements, we simulated images of \name\ using images of single
ultracool objects from other \HST/WFPC2 programs (GO-8563, PI
Kirkpatrick; GO-8581, PI Reid; and GO-8146, PI Reid). We only used
objects with an equivalent or higher signal-to-noise ratio (S/N)
compared to the science data so that we could degrade the S/N of the
single images to match the science data. We also restricted ourselves
to observations consisting of two or more images to allow robust
rejection of cosmic rays.  We only shifted images by an integer number
of pixels to preserve the somewhat undersampled nature of the WFPC2
data.  However, we were able to reproduce the binary configuration of
\name\ to within 0.2~pixels of the actual ($\Delta{x}$, $\Delta{y}$)
separation of ($+$3.2, $-$1.6)~pixels by carefully selecting
appropriate pairs of input PSFs whose sub-pixel positions were
determined in advance by single TinyTim PSF fitting.

We fitted the simulated binary images with TinyTim PSFs in the same
way as the science data. The resulting rms scatter of the
truth-minus-fitted parameters gave their errors, and the mean gave
their systematic offsets. For both the separation and P.A., the
offsets were small compared to the errors ($-$0.5$\pm$1.8~mas and
$-$0$\fdg$1$\pm$1$\fdg$1), but we found an offset in the flux ratio
that was large compared to its rms error ($-$0.10$\pm$0.02~mag). Since
these errors in deblending the binary are due to small imperfections
in the PSF model, it is natural that they would have the largest
impact on the flux ratio, not positional measurements, since the cores
of the PSFs are well-separated for this binary. In fact, because this
binary has a flux ratio near unity, the systematic offset we found
``flips'' the binary, changing the component that is identified as the
primary, and this flip brings the astrometry into agreement with the
astrometry from other epochs.

We applied the systematic offsets from our simulations to the binary
parameters, resulting in a separation of 161.1$\pm$1.8~mas, a P.A. of
57$\fdg$5$\pm$1$\fdg$1, and a flux ratio of 0.06$\pm$0.02~mag
(Table~\ref{tbl:astrom}). We can compare these parameters to those
derived by \citet{2003AJ....126.1526B}, who found a separation
(163.0$\pm$2.8~mas) and P.A. (57$\fdg$5$\pm$0$\fdg$3) consistent with
our measurements. Our separation uncertainty is slightly smaller, and
our P.A. uncertainty is larger. From their Figure~2, it is clear that
our separation and P.A. errors are actually consistent with their
analysis, and the apparent discrepancy between our errors only arises
because they condense their detailed study of binary parameter
uncertainties to a single number for the separation error and two
numbers for the P.A. error (0$\fdg$3 above separations of 150~mas;
1$\fdg$2 below 150~mas). For example, our P.A. error of 1$\fdg$1 is
intermediate between their two values, which is reasonable for a
binary with a separation very close to their cutoff between the two
regimes. However, the flux ratio of 0.36$\pm$0.07~mag derived by
\citet{2003AJ....126.1526B} is inconsistent with ours at 4$\sigma$.
Because the measurement of the flux ratio is most sensitive to
imperfections in the PSF model, it is natural that our different PSF
fitting methods would disagree most on this parameter. We note that
they apply a large 0.17~mag systematic offset to their flux ratio,
which again is a single number condensed from more detailed analysis
(see their Figure~3). This offset could account for 2.3$\sigma$ of the
discrepancy, which would bring our two flux ratios into reasonable
agreement. The F814W flux ratio does not enter substantially into the
following analysis, and so the discrepancy between our value and that
of \citet{2003AJ....126.1526B} has no impact on our results.

\tabletypesize{\small}
\begin{deluxetable*}{lccccc}
\tablecaption{Best-Fit Binary Parameters for \name \label{tbl:astrom}}
\tablewidth{0pt}
\tablehead{
\colhead{Epoch (UT)} &
\colhead{Instrument} &
\colhead{Filter} &
\colhead{$\rho$ (mas)} &
\colhead{P.A. (\degs)} &
\colhead{$\Delta{m}$ (mag)}}
\startdata
 2000 Aug 13  & \HST/WFPC2-PC1\tablenotemark{a}  &  F814W  &  161.1$\pm$1.8  & \phn57.5$\pm$1.1  &  0.06$\pm$0.02  \\
 2001 Sep 22  & Gemini/Hokupa`a\tablenotemark{a} &   \Kp\  &  167.7$\pm$1.0  & \phn68.2$\pm$0.5  &  0.04$\pm$0.02  \\
 2003 Jul 11  & VLT/NACO\tablenotemark{a}        &   $H$   &  164.2$\pm$0.3  & \phn89.2$\pm$0.2  &  0.05$\pm$0.05  \\
 2006 Jun 28  & VLT/NACO\tablenotemark{a}        &   \Ks\  &  131.3$\pm$0.3  &   130.38$\pm$0.16 & 0.077$\pm$0.017 \\
 2008 May 29  & Keck/NIRC2\tablenotemark{a}      &   \Ks\  & 119.32$\pm$0.14 &   170.07$\pm$0.11 & 0.067$\pm$0.010 \\
 2008 Sep  8  & Keck/NIRC2                       &   $J$   &  120.5$\pm$0.3  &   175.94$\pm$0.11 &  0.06$\pm$0.02  \\
              & Keck/NIRC2\tablenotemark{a}      &   $H$   &  120.2$\pm$0.3  &   176.05$\pm$0.10 & 0.065$\pm$0.017 \\
              & Keck/NIRC2                       &   \Ks\  &  120.4$\pm$0.4  &   175.89$\pm$0.11 &  0.06$\pm$0.02  \\
              & Keck/NIRC2                       &   \Lp\  &  121.1$\pm$0.5  &    176.1$\pm$0.4  & 0.004$\pm$0.016 \\
 2008 Dec  1  & Keck/NIRC2\tablenotemark{a}      &   \Ks\  &  121.4$\pm$1.0  &    180.9$\pm$0.8  &  0.06$\pm$0.06  \\
\enddata

\tablenotetext{a}{Used in the orbit fit.}

\end{deluxetable*}

\subsection{Gemini/Hokupa`a \label{sec:gem}}

2MASS~J2206$-$2047AB was imaged on UT 2001 September 22 by the
Hokupa`a curvature AO system at the Gemini-North Telescope on Mauna
Kea, Hawai`i. Analysis of these data has previously been presented by
\citet{2002ApJ...567L..53C}; however, we have conducted our own
analysis in an attempt to reduce the astrometric errors. We retrieved
these raw data from the Gemini science archive and registered,
sky-subtracted, and performed cosmic ray rejection on the images.
Figure~\ref{fig:data} shows a typical image from of one of the 15 \Kp\
band 10~s exposures which were used to derive the astrometry for
\name.

We used the same analytic PSF-fitting routine as for the Keck data to
fit the Gemini images of \name. Adopting an instrument pixel scale of
19.98$\pm$0.08~mas~\perpix\ (F. Rigaut 2001, private communication),
we found a separation of 167.7$\pm$1.0~mas, where the uncertainty is
the standard deviation of measurements from individual dithered
images. This is in good agreement with the 168$\pm$7~mas separation
reported by \citet{2002ApJ...567L..53C}. The improvement in the
separation error may be attributed to the fact that
\citet{2002ApJ...567L..53C} used the scatter among $J$, $H$, and
\Kp-band images, whereas we have restricted our measurement to the
bandpass with highest quality images (\Kp\ band). We found a \Kp\ band
flux ratio of 0.04$\pm$0.02~mag, which is consistent (at 1.1$\sigma$)
with the flux ratio of 0.08$\pm$0.03~mag derived by
\citet{2002ApJ...567L..53C}.

Unfortunately, we are not able to derive the correct value for the
binary P.A. from the archive images. As reported by
\citet{2002ApJ...567L..53C}, the image rotator was turned off for
these observations so that the pupil is aligned with the detector.
Thus, there is an arbitrary rotation in the images, which is not
recorded in the FITS headers, and this rotation changes during each
data set. We were able to remove this changing rotation by subtracting
the parallactic angle from the binary P.A. measured in the individual
dithers, and the rms scatter among the resulting P.A. measurements was
0$\fdg$4. Adding in quadrature the 0$\fdg$3 error in the absolute
orientation of Hokupa`a/QUIRC that was adopted by
\citet{2002ApJ...567L..53C} results in a P.A. error of 0$\fdg$5, which
is identical to their P.A. uncertainty.

Given the good agreement between our derived parameters and those
derived by \citet{2002ApJ...567L..53C}, we adopt our own separation
measurement because of the smaller uncertainty but the
\citet{2002ApJ...567L..53C} P.A. measurement because of our inability
to reconstruct the orientation of the archival images. As discussed in
Section~\ref{sec:orbit}, this choice is validated by the resulting
orbit fit having a reduced $\chi^2$ near unity.

\begin{figure}[b]
\plotone{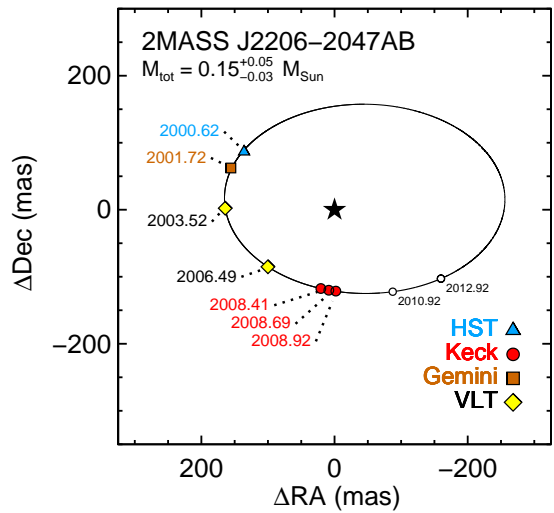}
\caption{\normalsize Relative astrometry along with the best fitting
  orbit for \name\ (reduced $\chi^2$ of 1.07 for 7 degrees of
  freedom). The empty circles show the predicted location of \nameB\
  in the future.  Error bars are smaller than the plotting
  symbols. The orbit is sufficiently well constrained that the
  uncertainty in the total mass is dominated by the 9.1\% parallax
  error. \label{fig:orbit-skyplot}}
\end{figure}

\begin{figure*}[t]
\plotone{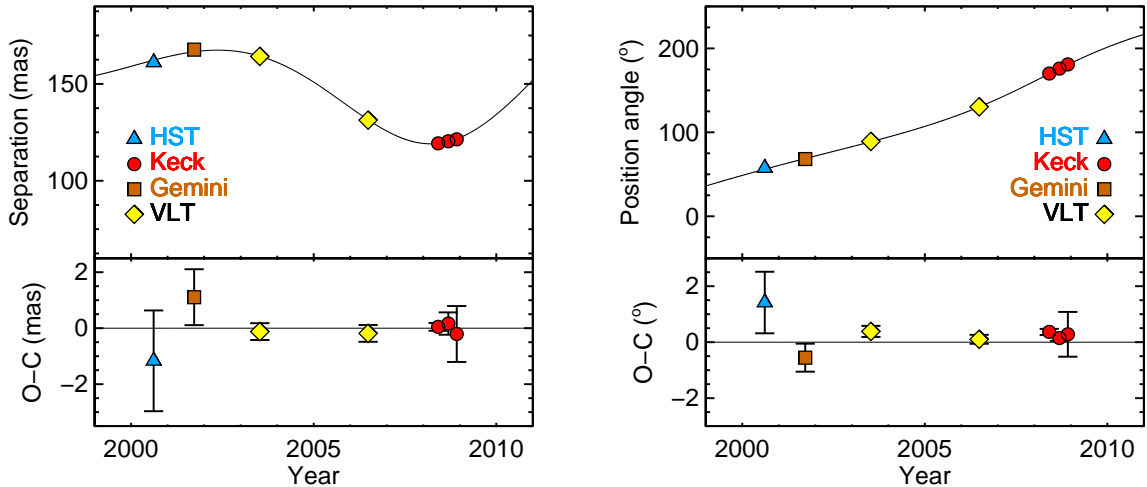}
\caption{\normalsize Measurements of the projected separation ({\em
    left}) and P.A. ({\em right}) of \name.  The best-fit orbit is
  shown as a solid line.  The bottom panels show the observed minus
  predicted measurements with observational error bars.
  \label{fig:orbit-sep-pa}}
\end{figure*}

\subsection{VLT/NACO \label{sec:vlt}}

We retrieved archival images of \name\ obtained with the VLT at
Paranal Observatory on UT 2003 July 11 and 2006 June 28. These data
were taken with the NACO adaptive optics system
\citep{2003SPIE.4841..944L, 2003SPIE.4839..140R} using the N90C10
dichroic and S13 camera
(13.221$\pm$0.017~mas~\perpix)\footnote[3]{\texttt{\url{http://www.eso.org/sci/facilities/paranal/instruments/ naco/doc/VLT-MAN-ESO-14200-2761\_v83.3.pdf}}}
at both epochs. We registered, sky-subtracted, and performed cosmic
ray rejection on the raw archival images. The 2003 data comprise four
$H$ band 60~s exposures, and the 2006 data comprise 14 \Ks\ band 60~s
exposures. Typical images from each data set are shown in
Figure~\ref{fig:data}.

We used the same analytic PSF-fitting routine as was used for the Keck
data to fit the VLT images, and the results are summarized in
Table~\ref{tbl:astrom}. The uncertainties were determined from the
standard deviation of measurements from individual dithers. Since no
PSF star was observed contemporaneously with \name, we were unable to
assess any additional systematic errors in the VLT astrometry by
fitting simulated binary images. However, in our previous work we have
found that the rms scatter among dithered VLT images has been a good
representation of errors derived from such simulations
\citep{me-2397a}.


\section{Results \label{sec:results}}

\subsection{Orbit Determination and Dynamical Mass \label{sec:orbit}}

Our observations together with archival data span 8.3~years of the
orbit of \name. We used a Markov Chain Monte Carlo (MCMC) approach,
described in detail by \citet{2008ApJ...689..436L}, to determine the
probability distributions of all orbital parameters. Chains all had
lengths of 2$\times$10$^8$ steps, and the correlation length of our
most correlated chain, as defined by \citet{2004PhRvD..69j3501T}, was
5.4$\times$10$^4$ for the orbital period. This gives an effective
length of the chain of 3.7$\times$10$^3$, which in turn gives
statistical uncertainties in the parameter errors of about
$1/\sqrt{3.7\times10^3}$~=~1.6\%, i.e., negligible.  The single
best-fit orbit has a reduced $\chi^2$ of 1.07 (7 degrees of freedom)
and is shown in Figures~\ref{fig:orbit-skyplot} and
\ref{fig:orbit-sep-pa}, and the best-fit parameters and their
confidence limits are given in Table~\ref{tbl:orbit}.

\begin{figure}
\plotone{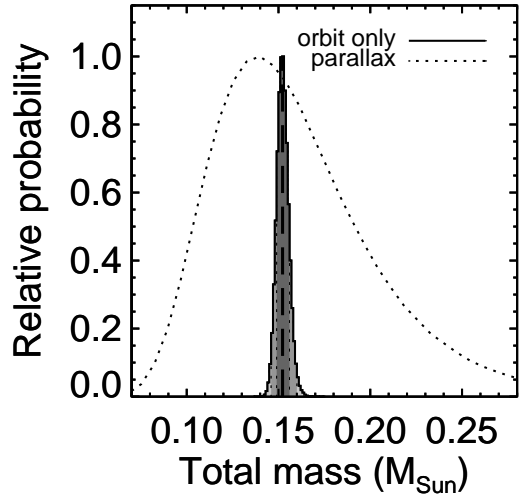}
\caption{\normalsize Probability distribution of the total mass of
  \name\ resulting from our MCMC analysis.  The histogram is shaded to
  indicate the 68.3\%, 95.4\%, and 99.7\% confidence regions, which
  correspond to 1$\sigma$, 2$\sigma$, and 3$\sigma$ for a normal
  distribution. The dashed line represents the median value of
  0.152~\Msun. The standard deviation of the distribution is
  0.003~\Msun. The dotted unshaded curve shows the final mass
  distribution after accounting for the additional $^{+32}_{-22}$\%
  error due to the uncertainty in the parallax.  The asymmetry in this
  curve is due to the asymmetric distance errors resulting from
  symmetric parallax errors.  The confidence limits for both
  distributions are given in
  Table~\ref{tbl:orbit}. \label{fig:orbit-mass}}
\end{figure}

\begin{figure*}
\plotone{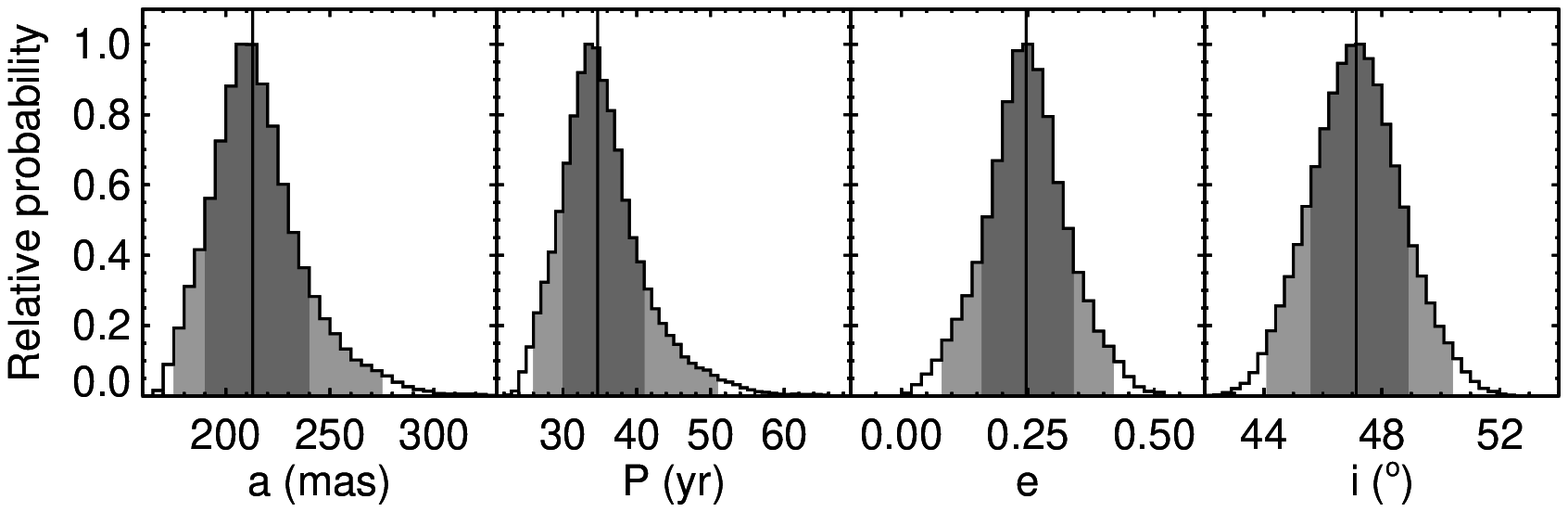} \\
\plotone{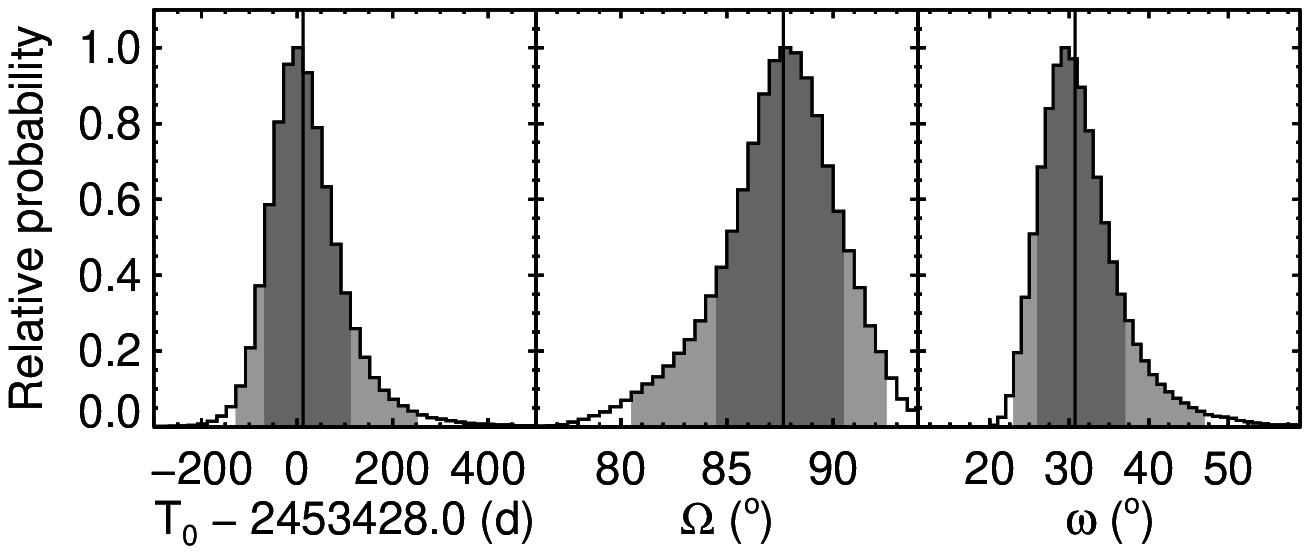}
\caption{\normalsize Probability distributions of all orbital
  parameters derived from the MCMC analysis: semimajor axis ($a$),
  orbital period ($P$), eccentricity ($e$), inclination ($i$), epoch
  of periastron ($T_0$), P.A. of the ascending node ($\Omega$), and
  argument of periastron ($\omega$).  Each histogram is shaded to
  indicate the 68.3\% and 95.4\% confidence regions, which correspond
  to 1$\sigma$ and 2$\sigma$ for a normal distribution, and the solid
  vertical lines represent the median values.  Note that $T_0$ is
  shown in days since UT 2005 March 11 00:00 for
  clarity. \label{fig:orbit-parms}}
\end{figure*}

\begin{figure}
\plotone{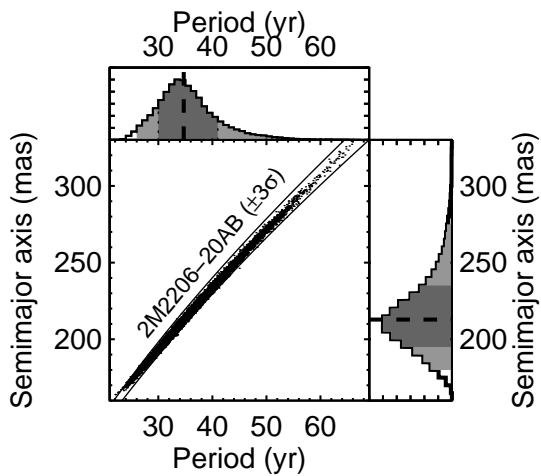}
\caption{\normalsize Steps in the MCMC chain show a high level of
  correlation between the orbital period and semimajor axis.  This
  correlation enables the total mass to be determined more precisely
  than from simple propagation of errors for these two parameters
  (\Mtot\ = $a^3/P^2$).  Lines are drawn demarcating the 3$\sigma$
  range for the total mass of \name\ without accounting for the
  distance uncertainty (0.143--0.163~\Msun). \label{fig:orbit-p-a}}
\end{figure}

Applying Kepler's Third Law ($\Mtot = a^3/P^2$) to the period and
semimajor axis distributions gives the posterior probability
distribution for the total mass of \name, which has a median of
0.152~\Msun, a standard deviation of 0.003~\Msun, and 68.3(95.4)\%
confidence limits of $^{+0.003}_{-0.003}$($^{+0.007}_{-0.006}$)~\Msun\
(Figure~\ref{fig:orbit-mass}). The resulting MCMC probability
distributions for all parameters are shown in
Figure~\ref{fig:orbit-parms}.  It is evident from
Figure~\ref{fig:orbit-p-a} that the tight correlation between the two
parameters $P$ and $a$ is responsible for the very precise total mass
(2\%), despite the fact the parameters are not independently
determined as precisely (16\% and 12\%, respectively). The MCMC
probability distribution of the total mass does not include the
uncertainty in the parallax (9.1\%), which by simple propagation of
errors would contribute an additional $^{+32}_{-22}$\% uncertainty in
the mass. We account for this additional error by randomly drawing a
normally distributed parallax value for each step in the chain. The
resulting mass distribution is asymmetric, and our final determination
of the total mass is 0.15$^{+0.05}_{-0.03}$($^{+0.13}_{-0.06}$)~\Msun\
at 68.3(95.4)\% confidence.

As an independent verification of our MCMC results, we also fit the
orbit of \name\ using the linearized least-squares routine \orbit\
\citep[described in][]{1999A&A...351..619F}. All of the orbital
parameters are consistent between the \orbit\ and MCMC results, and
the resulting total mass and $\chi^2$ were identical. Using \orbit, we
tested whether varying the input astrometry and corresponding
uncertainties affected the orbital solution. We tried a variety of
permutations: using the published \HST\ and/or Gemini astrometry
(rather than our own); using Keck astrometry with and without Monte
Carlo offsets and with rms or Monte Carlo-derived errors; and
excluding one of the two VLT epochs. The resulting orbits had masses
of 0.145--0.157~\Msun, with errors of 0.003--0.020~\Msun, and $\chi^2$
of 1.07--2.32. Thus, the dynamical mass and corresponding error are
not significantly impacted by the input astrometry, as all variation
is much smaller than the parallax error, and our default solution has
a lower $\chi^2$ and mass uncertainty than any other plausible trial
permutation.

\tabletypesize{\small}
\begin{deluxetable*}{lcccc}
\tablewidth{0pt}
\tablecaption{Derived Orbital Parameters for \name \label{tbl:orbit}}
\tablehead{
\colhead{}   &
\multicolumn{3}{c}{MCMC}      &
\colhead{\orbit\tablenotemark{a}} \\
\cline{2-4}
\colhead{Parameter}   &
\colhead{Median}      &
\colhead{68.3\% c.l.} &
\colhead{95.4\% c.l.} &
\colhead{} }
\startdata
Semimajor axis $a$ (mas)                                         &  213   &    $-$18, 24    &    $-$35, 60    &  \phs210$\pm$30\phn\phs\\
Orbital period $P$ (yr)                                          &   35   &     $-$5, 6     & \phn$-$9, 16    &   \phs35$\pm$8\phn \phs\\
Eccentricity $e$                                                 &  0.25  &  $-$0.08, 0.08  &  $-$0.17, 0.17  & \phs0.24$\pm$0.12  \phs\\
Inclination $i$ (\degs)                                          &  47.1  &   $-$1.5, 1.6   &     $-$3, 3     &   \phs47$\pm$2\phn \phs\\
Time of periastron passage $T_0-2453440.5$\tablenotemark{b} (JD) &    0   &    $-$60, 80    &   $-$130, 240   & \phn$-$1$\pm$80    \phs\\
P.A. of the ascending node $\Omega$ (\degs)                      &   88   &     $-$3, 2     &     $-$7, 5     &   \phs88$\pm$4\phn \phs\\
Argument of periastron $\omega$ (\degs)                          &   31   &     $-$4, 6     & \phn$-$7, 16    &   \phs31$\pm$6\phn \phs\\
Total mass (\Msun): fitted\tablenotemark{c}                      & 0.152  & $-$0.003, 0.003 & $-$0.006, 0.007 &\phs0.152$\pm$0.003 \phs\\
Total mass (\Msun): final\tablenotemark{d}                       &  0.15  &  $-$0.03, 0.05  &  $-$0.06, 0.13  & \phs0.15$\pm$0.04  \phs\\
Reduced $\chi^2$ (7 degrees of freedom)                          &  1.07  &     \nodata     &     \nodata     &          1.07      \phs\\
\enddata

\tablenotetext{a}{The orbital parameters determined by \orbit\ with
  their linearized 1$\sigma$ errors.}

\tablenotetext{b}{UT 2005 March 11 00:00:00.0} 

\tablenotetext{c}{The ``fitted'' total mass represents the direct
  results from fitting the observed orbital motion without accounting
  for the parallax error.  For the linearized \orbit\ error, the
  covariance between $P$ and $a$ was taken into account.}

\tablenotetext{d}{The ``final'' total mass includes the additional
  $^{+32}_{-22}$\% error in the mass due to the error in the
  parallax.}

\end{deluxetable*}

\subsection{Spectral Types \label{sec:spt}}

Using the integrated-light optical spectrum of \nameint, both
\citet{2000AJ....120.1085G} and \citet{2005A&A...441..653C} found a
spectral type of M8.0$\pm$0.5. Without resolved spectroscopy of the
binary, we cannot directly determine the spectral types of the two
components; however, our resolved photometry shows that they are
nearly identical. To quantify the potential difference in spectral
types, we compiled 2MASS photometry for the single M8.0, M8.5, and
M9.0 objects with parallaxes from \citet{1992AJ....103..638M} and
\citet{2002AJ....124.1170D}. Between the spectral types M8.0 and M8.5,
we found a difference in absolute magnitude of $\Delta{M_J}$ =
0.67$\pm$0.19~mag, $\Delta{M_H}$ = 0.68$\pm$0.20~mag, and
$\Delta{M_{K_S}}$ = 0.64$\pm$0.22~mag, where the uncertainty is the
rms of objects in each bin added in quadrature. Between spectral types
M8.0 and M9.0 we found only slightly larger magnitude differences. Our
best measurement of the flux ratio of \name\ in each of these bands is
$\Delta{J}$ = 0.06$\pm$0.02~mag, $\Delta{H}$ = 0.065$\pm$0.017~mag,
$\Delta{K_S}$ = 0.067$\pm$0.010~mag.\footnote[4]{Note that our flux
  ratios in the $J$ and $H$ bands are actually determined using MKO
  filters, but we neglect the slight differences in the 2MASS and MKO
  photometric systems for this comparison.}  Thus, we find that the
$J$, $H$, and $K_s$ band photometry is inconsistent with the two
components of \name\ having different spectral types at 3.0$\sigma$,
3.1$\sigma$, and 2.6$\sigma$, respectively.  Therefore, we adopt
spectral types of M8.0$\pm$0.5 for both components.

\begin{figure}
\plotone{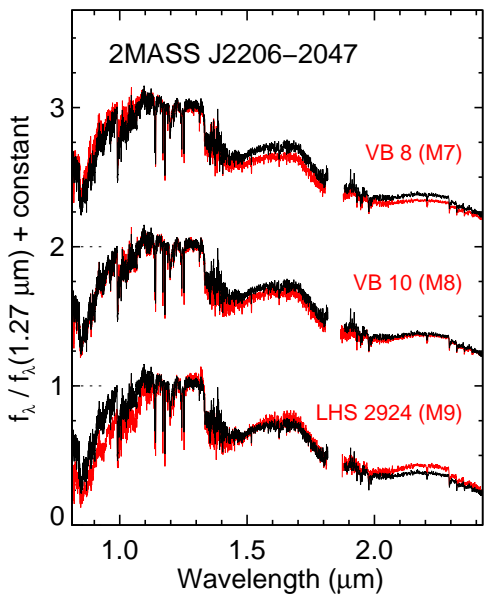}
\caption{\normalsize The integrated-light near-infrared spectrum of
  \nameint\ (black). Data for M~dwarf spectral standards (red) are
  shown for comparison \citep{2005ApJ...623.1115C}.  The best matching
  standard is VB 10 (M8), which is consistent with the optical
  spectral type of M8.0$\pm$0.5 for \nameint\
  \citep{2000AJ....120.1085G,
    2005A&A...441..653C}. \label{fig:spectra}}
\end{figure}

\subsection{Bolometric Luminosities \label{sec:lbol}}

Because of the identical spectral types and colors of the components
of \name, we computed their individual bolometric luminosities simply
from the flux ratio and integrated-light bolometric luminosity.  For
the former, we used the most precise flux ratio available, which was
0.067$\pm$0.010~mag from our 2008 September Keck $K_S$ band data.
There is no published value for the bolometric luminosity of \nameint,
so we computed it from its near-infrared spectrum and estimated \Lp\
band photometry.  We obtained the spectrum on UT 2008 July 6 using
IRTF/SpeX \citep{2003PASP..115..362R} in SXD mode, which has five
orders spanning 0.81--2.42~\micron\ ($R$~=~1200).  We calibrated,
extracted, and telluric-corrected the data using the SpeXtool software
package \citep{2003PASP..115..389V, 2004PASP..116..362C}.  In
Figure~\ref{fig:spectra} we show our spectrum alongside spectra from
\citet{2005ApJ...623.1115C} of the M~dwarf spectral standards defined
by \citet{1991ApJS...77..417K}.  As expected from the M8.0$\pm$0.5
optical spectral type of \nameint, the best matching standard is the
M8 dwarf VB~10.  To estimate the \Lp\ band photometry, we compiled the
$K-\Lp$ colors of M~dwarfs and early L~dwarfs from \citet{gol04},
which essentially follow a linear relation with spectral type.
Fitting all single M1--L1 objects (i.e., excluding the binaries LHS
333AB, LHS~2397aAB, and 2MASS~J0746+2000AB) and weighting by their
photometric errors, we found
\begin{eqnarray}
  K_{\rm MKO}-\Lp & = & 0.111~+~0.0526\times{\rm SpT}
\end{eqnarray}
where $K-\Lp$ is in mag, spectral type (SpT) is defined such that M0 =
0 and L0 = 10, and the rms about the fit was 0.07~mag.  From this
relation, we estimated a $K-\Lp$ color for \nameint\ of
0.53$\pm$0.07~mag, resulting in an \Lp\ band magnitude of
10.73$\pm$0.08~mag.\footnote[5]{We converted the 2MASS \Ks\ band
  integrated-light photometry to the MKO photometric system using
  synthetic photometry of our SpeX spectrum: $K_{\rm MKO}-K_{\rm
    2MASS}$ = $-$0.051~mag.}

To derive the integrated-light bolometric luminosity, we numerically
integrated our SpeX spectrum and the \Lp\ band photometric point at
3.8~\micron, interpolating between the gaps in the data, neglecting
flux at shorter wavelengths, and extrapolating beyond \Lp\ band
assuming a blackbody.  We determined the luminosity error in a Monte
Carlo fashion by adding randomly drawn noise to our data over many
trials and computing the rms of the resulting luminosities. We
accounted both for the noise in the spectrum and the errors in the
2MASS photometry used to flux-calibrate it.  Before accounting for the
error in the distance we found a total bolometric luminosity of
$\log(\Lbol/\Lsun)$ = $-$2.982$\pm$0.006~dex.  After accounting for
this error, the symmetric parallax uncertainty results in slightly
asymmetric luminosity errors, giving $\log(\Lbol/\Lsun)$ =
$-$2.98$^{+0.08}_{-0.07}$~dex.  Using the flux ratio to apportion this
luminosity to the two binary components results in individual
luminosities of $-$3.27$^{+0.08}_{-0.07}$~dex and
$-$3.30$^{+0.08}_{-0.07}$~dex.  In the following analysis, we
correctly account for the covariance between these quantities (via the
flux ratio), enabling more precise determinations of relative
quantities (such as $\Delta\Teff$) due to the more precise luminosity
ratio ($\Delta\log(\Lbol)$ = 0.027$\pm$0.004~dex).

\subsection{Atmospheric Model Fitting: \Teff, \logg, and $R$ \label{sec:atm-fit}}

Because the two components of \name\ have essentially identical fluxes
and colors, we can determine the effective temperatures and surface
gravities of both by fitting atmospheric models to its
integrated-light spectrum.\footnote{In Section~\ref{sec:teff-logg}, we
  use the measured total mass and luminosity ratio to derive from
  evolutionary models an effective temperature difference of
  27$\pm$5~K between the two components.  Since the model grid steps
  are 100~K, a single-temperature fit to the integrated-light spectrum
  is valid.}  We used the PHOENIX-Gaia \citep{2005ESASP.576..565B} and
the Ames-Dusty \citep{2001ApJ...556..357A} solar-metallicity
atmospheric models to fit our IRTF/SpeX SXD spectrum of \nameint.  The
PHOENIX-Gaia models include updated line lists compared to the
AMES-Dusty models, but they do not include the effects of dust.  The
treatment of dust in the Ames-Dusty models is an extreme limiting case
(no dust settling), but models with a more sophisticated treatment of
dust are not yet publicly available.  For the PHOENIX-Gaia models, we
used grids of synthetic spectra ranging in \Teff\ from 2000 to 3500~K
($\Delta$\Teff~=~100~K) and \logg\ from 3.5 to 5.5
($\Delta$\logg~=~0.5).  For the Ames-Dusty models, we used grids of
synthetic spectra ranging in \Teff\ from 1500 to 3400~K
($\Delta$\Teff~=~100~K) and \logg\ from 4.0 to 6.0
($\Delta$\logg~=~0.5).

\begin{figure}[h]
\includegraphics[height=3.4in,angle=90]{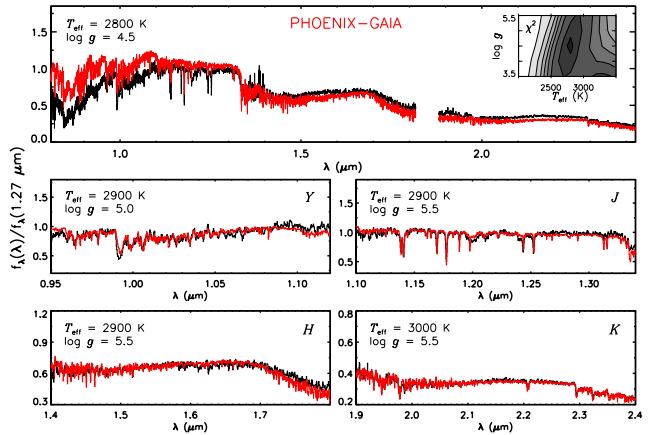}
\caption{ \normalsize The integrated-light near-infrared spectrum of
  \nameint\ (black) shown with the best fitting PHOENIX-Gaia models
  (red) for the entire spectrum (top) and for individual spectral
  ranges ($Y$, $J$, $H$, and $K$ bands). The top panel inset (upper
  right) shows $\chi^2$ contours of 1.02, 1.1, 1.3, 1.5, 1.7, 2.0,
  2.3, and 3.3 times the minimum $\chi^2$ for model atmospheres
  compared to the entire spectrum.  The model fit is well-constrained
  in \Teff\ and less so in \logg. \label{fig:gaia-fit}}
\end{figure}

\begin{figure}[h]
\includegraphics[height=3.4in,angle=90]{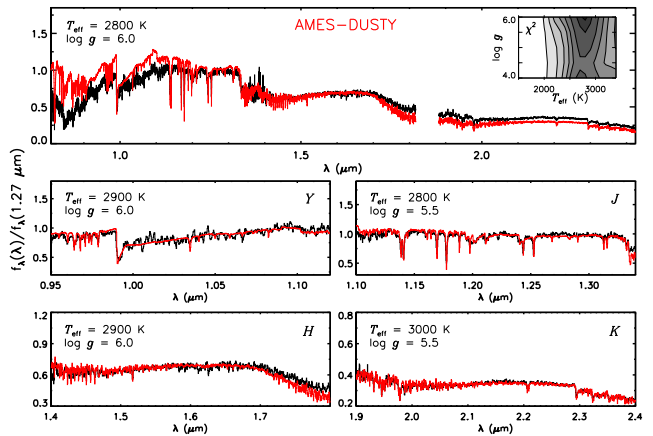}
\caption{ \normalsize The integrated-light near-infrared spectrum of
  \nameint\ (black) shown with the best fitting Ames-Dusty models
  (red) for the entire spectrum (top) and for individual spectral
  ranges ($Y$, $J$, $H$, and $K$ bands). The top panel inset (upper
  right) shows $\chi^2$ contours of 1.02, 1.1, 1.3, 1.5, 1.7, 2.0,
  2.3, and 3.3 times the minimum $\chi^2$ for model atmospheres
  compared to the entire spectrum.  The model fit is well-constrained
  in \Teff\ and that the minimum $\chi^2$ is only achieved when the
  model \logg\ reaches the upper limit of the
  grid. \label{fig:dusty-fit}}
\end{figure}

Our fitting procedure utilized a Monte Carlo approach based on that of
\citet{2008ApJ...678.1372C} and Bowler, Liu \& Cushing (2009,
submitted).  To account for the heterogeneous resolution of our SXD
spectrum, we Gaussian smoothed synthetic spectra in separate spectral
ranges corresponding to the different SXD orders.  When fitting our
near-infrared spectrum (0.81--2.42~\micron), we excluded a small
region from 1.82--1.88~\micron\ not covered by the instrument.  We
flux-calibrated our observed spectrum using 2MASS $J$, $H$, and \Ks\
photometry.  For each Monte Carlo trial, we applied small flux shifts
to the observed spectrum corresponding to the spectroscopic (SpeX) and
photometric (2MASS) measurement errors and then found the best fitting
model by minimizing the $\chi^2$ statistic.  This process was repeated
10$^3$ times, after which we tallied the fraction of times each model
yielded the best fit ($f_{\rm MC}$).  Fractions near 1.0 indicate that
only a single model fit the data well.  The results of our fitting
procedure are given in Table~\ref{tbl:atm-fit}, and the best-fit
spectra for each set of models are shown in Figures~\ref{fig:gaia-fit}
and \ref{fig:dusty-fit}.

Both sets of models gave best-fit effective temperatures of 2800~K,
but the best-fit PHOENIX-Gaia model had a lower surface gravity
(\logg~=~4.5) than the Ames-Dusty model (\logg~=~6.0; the grid
maximum).  Although the $f_{\rm MC}$ value for the best-fit
PHOENIX-Gaia model was only 0.46, the next best-fit model ($f_{\rm
  MC}$~=~0.27) had a temperature and gravity different by only one
grid step (\Teff~=~2700~K, \logg~=~4.0).  We also fitted the observed
spectrum separately over individual bandpasses ($Y$, $J$, $H$, and
$K$), and these fits yielded similar results to the entire spectrum,
typically within 100~K and 0.5 dex (i.e., one model grid step).  Thus,
we adopt errors of $\pm$100~K and $\pm$0.5~dex on the best-fit
parameters in order to account for the impact of the measurement
errors on the fit as well as uncertainties in modeling a limited
spectral range.

In addition to the effective temperature and surface gravity, the
radius can be derived from atmospheric model fitting when the distance
is known. This is because the scaling factor used to shift the
synthetic spectrum to the observed flux-calibrated spectrum is a free
parameter equal to $R^2/d^2$. Accounting for the flux ratio between
the two components, the error in the distance, and the rms scatter in
this scaling factor over the 10$^3$ trials, we found identical radii
of 0.096$\pm$0.009~\Rsun\ from the PHOENIX-Gaia models and
0.095$\pm$0.009~\Rsun\ from the Ames-Dusty models.\footnote{Because
  more than one PHOENIX-Gaia model fit the data well (see
  Table~\ref{tbl:atm-fit}), we used only the best fitting model's
  scaling factors when deriving radii.}

\tabletypesize{\small}
\begin{deluxetable}{rccc}
\tablewidth{0pt}
\tablecaption{ Best-fit Atmospheric Models of \name \label{tbl:atm-fit}}
\tablehead{
\colhead{Spectral range}    &
\colhead{\Teff\ (K)}      &
\colhead{\logg\ } &
\colhead{$f_{\rm MC}$\tablenotemark{a}} }
\startdata

\multicolumn{4}{c}{\bf PHOENIX-Gaia \citep{2005ESASP.576..565B}} \\
\cline{1-4}
All (0.81--2.42~\micron)  &  2800                   &  4.5                   &  0.46  \\
                          &  2700                   &  4.0                   &  0.27  \\
                          &  3400                   &  3.5\tablenotemark{b}  &  0.19  \\
$Y$ (0.95--1.12~\micron)  &  2900                   &  5.0                   &  1.00  \\
$J$ (1.10--1.34~\micron)  &  2900                   &  5.5\tablenotemark{b}  &  0.99  \\
$H$ (1.40--1.80~\micron)  &  2900                   &  5.5\tablenotemark{b}  &  1.00  \\
$K$ (1.90--2.40~\micron)  &  3000                   &  5.5\tablenotemark{b}  &  1.00  \\
\cline{1-4}

\multicolumn{4}{c}{} \\
\multicolumn{4}{c}{\bf Ames-Dusty \citep{2001ApJ...556..357A}} \\
\cline{1-4}
All (0.81--2.42~\micron)  &  2800                   &  6.0\tablenotemark{b}  &  0.99  \\
$Y$ (0.95--1.12~\micron)  &  2900                   &  6.0\tablenotemark{b}  &  0.54  \\
                          &  2800                   &  5.5                   &  0.46  \\
$J$ (1.10--1.34~\micron)  &  2800                   &  5.5                   &  1.00  \\
$H$ (1.40--1.80~\micron)  &  2900                   &  6.0\tablenotemark{b}  &  0.95  \\
$K$ (1.90--2.40~\micron)  &  3000                   &  5.5                   &  0.63  \\
                          &  3000                   &  6.0\tablenotemark{b}  &  0.37  \\

\enddata

\tablenotetext{a}{Fraction of Monte Carlo trials in which the model
  gave the best fit.}

\tablenotetext{b}{The best-fit value is at the edge of the model
  grid.}

\end{deluxetable}

\subsection{Age Constraints from Kinematics and Activity
  \label{sec:age}}

In this section, we consider whether the space motion or activity of
\nameint\ can provide useful constraints on the age of the system.
There are three values of its radial velocity in the literature: (1)
16.3$\pm$2.7~\kms\ derived by \citet{2002AJ....124..519R} from
cross-correlation of the optical spectrum with radial velocity
standards; (2) 8.0$\pm$2.0~\kms\ also from \citet{2002AJ....124..519R}
but derived from the central wavelength of the H$\alpha$ emission
line; and (3) 10.8$\pm$1.3~\kms\ derived by
\citet{2003A&A...401..677G} using the central wavelengths of
unspecified spectral lines.  \citet{2002AJ....124..519R} attributed
the discrepancy between their two radial velocities to the fact that
\nameint\ is a fast rotator ($v\sin{i}$ = 22~\kms) with an asymmetric
H$\alpha$ profile, confusing their estimate of the H$\alpha$ centroid.
\nameint's binarity was unknown to \citet{2002AJ....124..519R}, and we
note that this could be partially responsible for the discrepancy.
For example, if one component dominated the H$\alpha$ emission, the
H$\alpha$ centroid would be offset from the cross-correlation velocity
(which likely represents the average velocity of the two components)
by 1.8~\kms, assuming a mass ratio of unity.  However, this alone is
insufficient to account for the 8.3~\kms\ discrepancy observed.

\begin{figure*}
\plotone{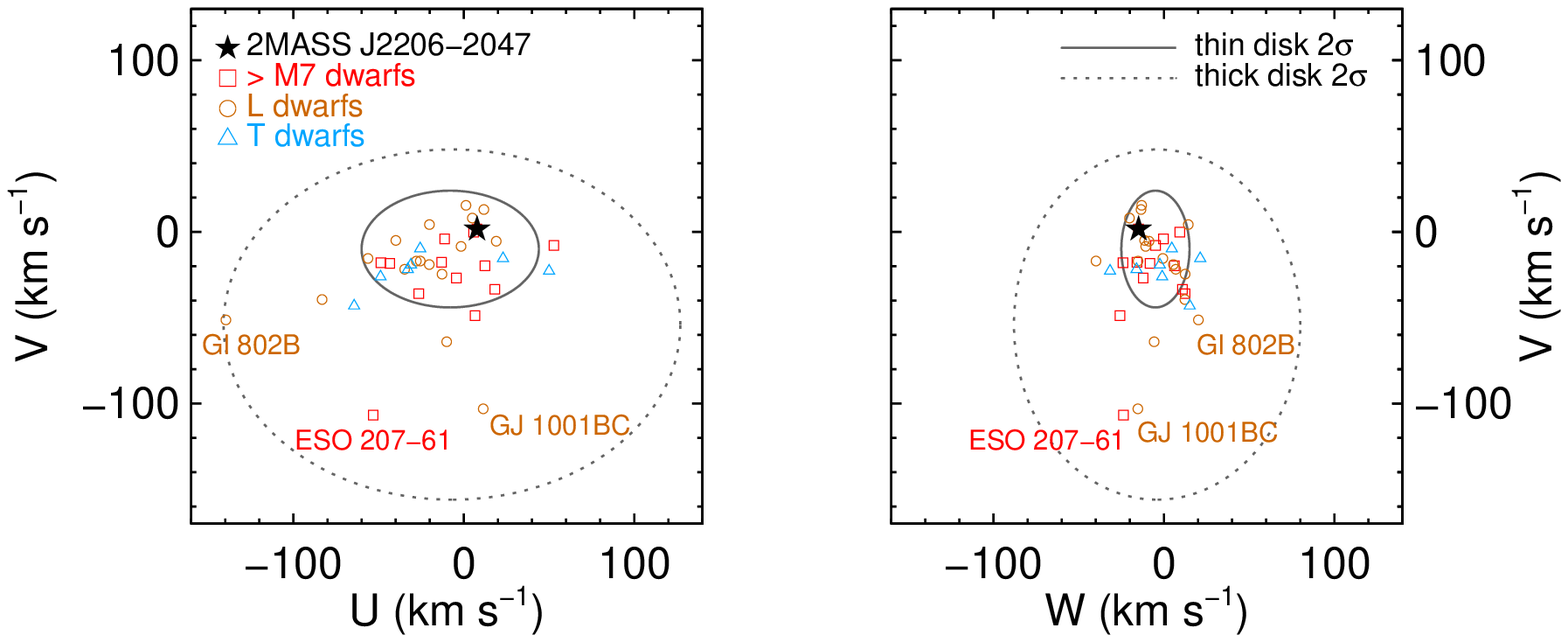}
\caption{ \normalsize The heliocentric space velocity of \nameint\
  (star) shown alongside other ultracool dwarfs: $>$~M7 dwarfs
  (squares), L dwarfs (circles), and T dwarfs (triangles). The
  2$\sigma$ ellipsoids of the thin disk (solid line) and thick disk
  (dotted line) as predicted by the Besan\c{c}on galaxy model
  \citep{2003A&A...409..523R} are also shown for comparison. The space
  velocity of \nameint\ is consistent with other ultracool dwarfs, and
  we derive a $>$~99.9\% thin disk membership
  probability. \label{fig:uvw}}
\end{figure*}

We used the cross-correlation radial velocity (16.3$\pm$2.7~\kms) from
\citet{2002AJ....124..519R} and the parallax and proper motion from
\citet{2006AJ....132.1234C} to derive the heliocentric velocity of
\nameint: $(U, V, W)$~=~($+$7.8$\pm$1.6, $+$1.7$\pm$1.1,
$-$15.0$\pm$2.1)~\kms.  We adopted the sign convention for $U$ that is
positive toward the Galactic center and accounted for the errors in
the parallax, proper motion, and radial velocity in a Monte Carlo
fashion.  For comparison, we compiled all objects of spectral type M7
or later that have the radial velocities, parallaxes, and proper
motions necessary for computing space motions \citep[described in
detail in Section~3.4 of][]{me-2397a}.  \nameint\ is only 1.3$\sigma$
away from the mean of this population's space motion ellipsoid
(Figure~\ref{fig:uvw}). Thus, its space motion is not significantly
different from other ultracool dwarfs, implying an age consistent with
the population of ultracool dwarfs as a whole.  Several authors have
attempted to estimate the age of this population, typically comparing
the distribution of tangential velocities ($V_{\rm tan}$, which
requires only a proper motion and distance determination) to the well
studied nearby populations of FGKM stars.  The resulting age for the
population of ultracool dwarfs estimated in this way has been found to
be 2--4~Gyr \citep{2002AJ....124.1170D,
  2009AJ....137....1F}.\footnote{\citet{2007ApJ...666.1205Z}
  determined a somewhat younger age ($\sim$1~Gyr) for the population
  of ultracool dwarfs, based on the small sample of L and T~dwarfs
  will full space velocities (21 objects). However, since L and
  T~dwarfs span a wider range of masses than earlier type objects, a
  typical IMF that rises at lower masses will naturally increase the
  number of young objects in this sample, biasing a kinematically
  derived age \citep[e.g., see Section~4.5
  of][]{2002AJ....124.1170D}.}

We have also assessed \nameint's membership in the Galactic
populations of the thin disk \citep[1--10~Gyr;
e.g.,][]{1998ApJ...497..870W} and thick disk \citep[$\sim$10~Gyr;
e.g.,][]{2002A&A...394..927I} using the Besan\c{c}on model of the
Galaxy \citep{2003A&A...409..523R}.  Our method is described in
Section~3.4 of our study of \lhsint\ \citep{me-2397a}, and for
\nameint\ we found a membership probability of $>$~99.9\% for the
thin disk and $<$~0.1\% for the thick disk.

Finally, the fact that \nameint\ is chromospherically active
\citep[$\log(L_{\rm H\alpha}/L_{\rm bol})$~=~$-$4.59,
$-$4.54;][respectively]{2000AJ....120.1085G, 2002AJ....124..519R}
could also potentially provide an age constraint, as the activity of
M~dwarfs changes with age. \citet{2008AJ....135..785W} showed that the
fraction of active M~dwarfs as a function the vertical distance above
the Galactic plane ($z$) provides a constraint on the activity
lifetime of M~dwarfs, given a model of how thick disk heating pumps up
$z$ over time.  \citet{2008AJ....135..785W} found that the activity
lifetime increases monotonically with M~dwarf spectral type, and the
latest type for which they were able to determine a robust lifetime
was M7 (8.0$^{+0.5}_{-1.0}$~Gyr). This provides a weak constraint on
the age of \nameint, as its activity is therefore expected to last for
at least $\gtrsim$8~Gyr.


\section{Tests of Models \label{sec:tests}}

Our measured total mass of \name\ enables strong tests of theoretical
models, and in the following analysis we consider two independent sets
of evolutionary models: the Tucson models \citep{1997ApJ...491..856B}
and the Lyon Dusty models \citep{2000ApJ...542..464C}.  Our approach
follows previous work for \twomassbin\ \citep{2008ApJ...689..436L},
\hdbin\ \citep{2009ApJ...692..729D}, and \lhsbin\ \citep{me-2397a}.
We drew measured properties (summarized in Table~\ref{tbl:meas}) from
random distributions, carefully accounting for the covariance between
different quantities (e.g., \Mtot\ and \Lbol\ are correlated through
the distance), and we used \Lbol\ (rather than \Teff) as the basis for
our model comparisons.

\subsection{Model-Inferred Age \label{sec:modelage}}

As described in detail by \citet{2008ApJ...689..436L} and
\citet{2009ApJ...692..729D}, the total mass of a binary along with its
individual component luminosities can be used to estimate the age of
the system from evolutionary models. This age estimate can be
surprisingly precise when both components are likely to be substellar
since their luminosities depend very sensitively on age. However, with
spectral types of M8.0$\pm$0.5, both components of \name\ are likely
to be stars unless the system is quite young.

\begin{figure}[b]
\includegraphics[width=3.4in]{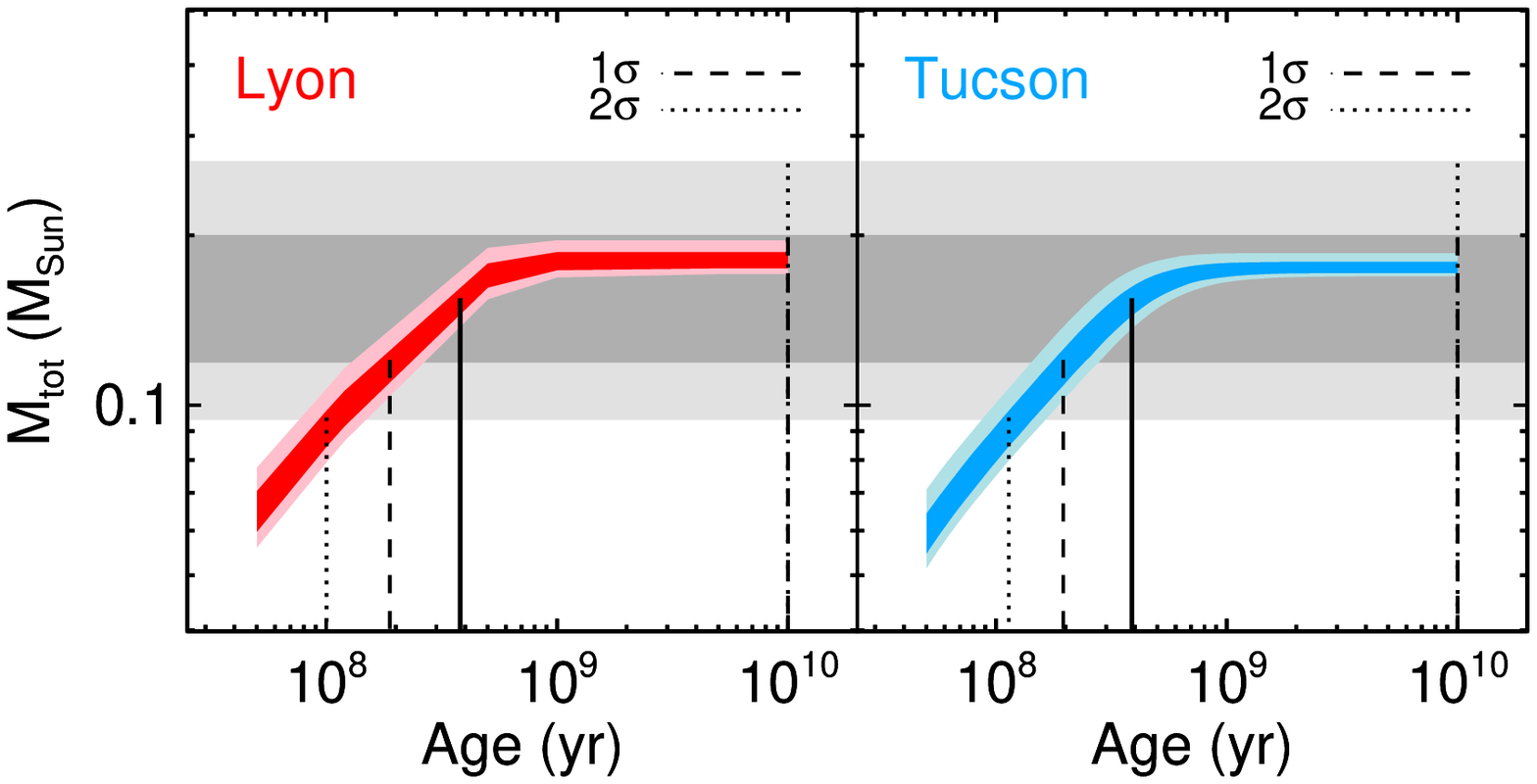}
\caption{ \normalsize Total mass (\Mtot) predicted by evolutionary
  models as a function of age, given the observational constraint of
  the luminosities of the two components of \name. The curved shaded
  regions show the 1$\sigma$ and 2$\sigma$ ranges in this
  model-derived mass. By applying the additional constraint of the
  measured total mass (\Mtot), we used the models to determine the age
  of \name\ (see Section~\ref{sec:modelage}). The horizontal gray bars
  show our 1$\sigma$ and 2$\sigma$ constraints on the total mass, and
  the resulting median, 1$\sigma$, and 2$\sigma$ model-inferred ages
  are shown by solid, dashed, and dotted lines, respectively.
  Model-inferred ages are truncated at 10~Gyr (the oldest age included
  in both sets of models), which happens at 1$\sigma$ for both models.
  Note that the upper limit of the \Mtot\ distribution corresponds to
  both components of \name\ being stars, and evolutionary models do
  not reach faint enough luminosities for such high mass objects at
  any age. \label{fig:mtot-age}}
\end{figure}

We derived an age of 0.4$^{+9.6}_{-0.2}$~Gyr from both the Tucson and
Lyon models (Figure~\ref{fig:mtot-age}).  Because the median total
mass is 0.15~\Msun, which is roughly the limit at which both
components would be brown dwarfs, the median age derived from models
is correspondingly young.  However, the 1$\sigma$ uncertainty in the
total mass reaches 0.20~\Msun, in which case both components would be
main-sequence stars.  In this case, since stars do not dim over time
as brown dwarfs do, the luminosities of both components do not
strongly constrain the age of the system.  Thus, while the lower bound
of our uncertainty on the model-derived age corresponds to the age a
pair of brown dwarfs would need to be to match the observed
luminosities and total mass, the upper limit is essentially
unconstrained.  Our limit of 10~Gyr comes from the fact that the
evolutionary models are computed only up to this age.  In our
analysis, which uses a Monte Carlo approach to compute model-derived
properties, we found that about 30\% of the time the randomly drawn
observed luminosities were too low to match the randomly drawn total
mass.  In other words, models would never predict that such massive
objects ($\gtrsim$0.09~\Msun\ stars) could be as faint as the
components of \name, and in such cases we assigned an age of 10~Gyr.

\tabletypesize{\footnotesize}
\begin{deluxetable}{lccc}
\tablewidth{0pt}
\tablecaption{Measured Properties of \name \label{tbl:meas}}
\tablehead{
\colhead{Property\tablenotemark{a}}   &
\colhead{Component A} &
\colhead{Component B} &
\colhead{Ref.}       }
\startdata
\Mtot\ (\Msun)             &        \multicolumn{2}{c}{    0.15$^{+0.05}_{-0.03}$  }         &   1   \\
Semimajor axis (AU)        &        \multicolumn{2}{c}{     5.8$^{+0.8}_{-0.7}$    }         &  1,2  \\
$d$ (pc)                   &        \multicolumn{2}{c}{    26.7$^{+2.6\phn}_{-2.1\phn}$    }         &   2   \\
Spectral type              &\phs       M8.0$\pm$0.5\phn   &\phs       M8.0$\pm$0.5\phn        &  1,3  \\
\Teff\ (K)                 &\phs \phn2800$\pm$100\tablenotemark{b} &\phs \phn2800$\pm$100\tablenotemark{b} &  1   \\
$J$ (mag)                  &\phs      13.07$\pm$0.02  &\phs      13.13$\pm$0.02      &  1,4  \\
$H$ (mag)                  &\phs      12.44$\pm$0.02  &\phs      12.51$\pm$0.02      &  1,4  \\
$K$ (mag)                  &\phs      11.98$\pm$0.03  &\phs      12.05$\pm$0.03      &  1,4  \\
$J-H$ (mag)                &\phs\phn   0.63$\pm$0.03  &\phs\phn   0.62$\pm$0.03      &  1,4  \\
$H-K$ (mag)                &\phs\phn   0.46$\pm$0.04  &\phs\phn   0.45$\pm$0.04      &  1,4  \\
$J-K$ (mag)                &\phs\phn   1.08$\pm$0.04  &\phs\phn   1.08$\pm$0.04      &  1,4  \\
$M_J$ (mag)                &\phs      10.93$\pm$0.19  &\phs      10.99$^{+0.19\phn}_{-0.20\phn}$      & 1,2,4 \\
$M_H$ (mag)                &\phs      10.31$^{+0.18\phn}_{-0.20\phn}$  &\phs      10.38$^{+0.18\phn}_{-0.20\phn}$      & 1,2,4 \\
$M_K$ (mag)                &\phs\phn   9.86$^{+0.18\phn}_{-0.20\phn}$  &\phs\phn   9.92$^{+0.18\phn}_{-0.20\phn}$      & 1,2,4 \\
$\log$(\Lbol/\Lsun)        &\phn    $-$3.27$^{+0.08\phn}_{-0.07\phn}$  &\phn     $-$3.30$^{+0.08\phn}_{-0.07\phn}$     &  1,2  \\
$\Delta\log$(\Lbol)        &          \multicolumn{2}{c}{ 0.027$\pm$0.004 }          &   1   \\
\enddata

\tablenotetext{a}{All near-infrared photometry on the MKO system.}

\tablenotetext{b}{Based on spectral synthesis model fitting.}

\tablerefs{(1)~This work; (2)~\citet{2006AJ....132.1234C};
  (3)~\citet{2000AJ....120.1085G, 2005A&A...441..653C}; (4)~\citet{2mass}.}

\end{deluxetable}

\tabletypesize{\footnotesize}
\begin{deluxetable}{lccc}
\tablewidth{0pt}
\tablecaption{Evolutionary Model-derived Properties of \name \label{tbl:model}}
\tablehead{
\colhead{Property}    &
\colhead{Median}      &
\colhead{68.3\% c.l.} &
\colhead{95.4\% c.l.} }
\startdata
                           \multicolumn{4}{c}{\bf Tucson Models \citep{1997ApJ...491..856B}} \\
                           \cline{1-4}
                           \multicolumn{4}{c}{System} \\
                           \cline{1-4}
Age (Gyr)\tablenotemark{a}    & 0.4     & $-    0.2, 9.6    $ & $-    0.3, 9.6    $ \\
$q$ ($M_{\rm B}/M_{\rm A}$)   & 0.981   & $-  0.007, 0.012  $ & $-  0.013, 0.014  $ \\
$\Delta$\Teff\ (K)            & 27      & $-      5, 5      $ & $-      9, 11     $ \\
                           \cline{1-4}
                           \multicolumn{4}{c}{Component A} \\
                           \cline{1-4}
$M_{\rm A}$ (\Msun)           & 0.077   & $-  0.017, 0.010  $ & $-  0.030, 0.014  $ \\
$T_{\rm eff,A}$ (K)           & 2660    & $-    100, 90     $ & $-    190, 170    $ \\
$\log(g_{\rm A})$ (cgs)       & 5.26    & $-   0.17, 0.09   $ & $-   0.33, 0.12   $ \\
$R_{\rm A}$ (\Rsun)           & 0.109   & $-  0.006, 0.008  $ & $-  0.010, 0.018  $ \\
Li$_{\rm A}$/Li$_0$           & 0.0    & $-    0.0, 0.8    $ & $-    0.0, 1.0    $ \\
                           \cline{1-4}
                           \multicolumn{4}{c}{Component B} \\
                           \cline{1-4}
$M_{\rm B}$ (\Msun)           & 0.076   & $-  0.017, 0.011  $ & $-  0.029, 0.015  $ \\
$T_{\rm eff,B}$ (K)           & 2640    & $-    100, 90     $ & $-    200, 170    $ \\
$\log(g_{\rm B})$ (cgs)       & 5.26    & $-   0.17, 0.10   $ & $-   0.33, 0.13   $ \\
$R_{\rm B}$ (\Rsun)           & 0.108   & $-  0.006, 0.009  $ & $-  0.010, 0.018  $ \\
Li$_{\rm B}$/Li$_0$           & 0.0    & $-    0.0, 0.9    $ & $-    0.0, 1.0    $ \\
                           \cline{1-4}
                           \multicolumn{4}{c}{} \\
                           \multicolumn{4}{c}{\bf Lyon Models \citep[Dusty;][]{2000ApJ...542..464C}} \\
                           \cline{1-4}
                           \multicolumn{4}{c}{System} \\
                           \cline{1-4}
Age (Gyr)\tablenotemark{a}    & 0.4     & $-    0.2, 9.6    $ & $-    0.3, 9.6    $ \\
$q$ ($M_{\rm B}/M_{\rm A}$)   & 0.982   & $-  0.006, 0.008  $ & $-  0.012, 0.011  $ \\
$\Delta$\Teff\ (K)            & 27      & $-      4, 5      $ & $-      8, 9      $ \\
                           \cline{1-4}
                           \multicolumn{4}{c}{Component A} \\
                           \cline{1-4}
$M_{\rm A}$ (\Msun)           & 0.077   & $-  0.017, 0.012  $ & $-  0.030, 0.018  $ \\
$T_{\rm eff,A}$ (K)           & 2550    & $-    100, 90     $ & $-    200, 170    $ \\
$\log(g_{\rm A})$ (cgs)       & 5.17    & $-   0.18, 0.12   $ & $-   0.34, 0.13   $ \\
$R_{\rm A}$ (\Rsun)           & 0.119   & $-  0.007, 0.010  $ & $-  0.011, 0.021  $ \\
Li$_{\rm A}$/Li$_0$           & 0.2    & $-   0.2, 0.5   $ & $-   0.2, 0.8   $ \\
                           \cline{1-4}
                           \multicolumn{4}{c}{Component B} \\
                           \cline{1-4}
$M_{\rm B}$ (\Msun)           & 0.076   & $-  0.017, 0.013  $ & $-  0.029, 0.018  $ \\
$T_{\rm eff,B}$ (K)           & 2530    & $-    100, 90     $ & $-    200, 170    $ \\
$\log(g_{\rm B})$ (cgs)       & 5.17    & $-   0.18, 0.12   $ & $-   0.34, 0.14   $ \\
$R_{\rm B}$ (\Rsun)           & 0.118   & $-  0.007, 0.010  $ & $-  0.011, 0.021  $ \\
Li$_{\rm B}$/Li$_0$           & 0.2    & $-   0.2, 0.5   $ & $-   0.2, 0.8   $ \\
\enddata

\tablenotetext{a}{Both sets of evolutionary models are only computed
  up to an age of 10~Gyr; therefore, this defines the upper limit on
  the model-derived ages.}

\end{deluxetable}

\subsection{Individual Masses \label{sec:qratio}}

Given the near unity flux ratio of \name, we expect the mass ratio to
also be very close to unity. We used evolutionary models to estimate
the mass ratio of \name\ ($q$~$\equiv$~$M_{\rm B}/M_{\rm A}$) by
constraining the model-derived individual masses of \nameA\ and
\nameB\ to add up to the observed total mass, while still matching
their observed luminosities. The Tucson models gave
$q$~=~0.981$^{+0.012}_{-0.007}$, while the Lyon models gave a
consistent value of 0.982$^{+0.008}_{-0.006}$. The resulting
individual masses (Table~\ref{tbl:model}) are essentially identical to
those resulting from an assumed mass ratio of unity, with the
exception of the upper confidence limits. This is due to the effect
described in Section~\ref{sec:modelage} where about 30\% of the
randomly drawn total masses and individual luminosities were
inconsistent with any models. In these cases, we assigned the highest
individual masses for which the luminosities were consistent with the
models. In Section~\ref{sec:add-mass}, we consider the issue of
plausible individual masses in more detail.

\begin{figure*}
\plotone{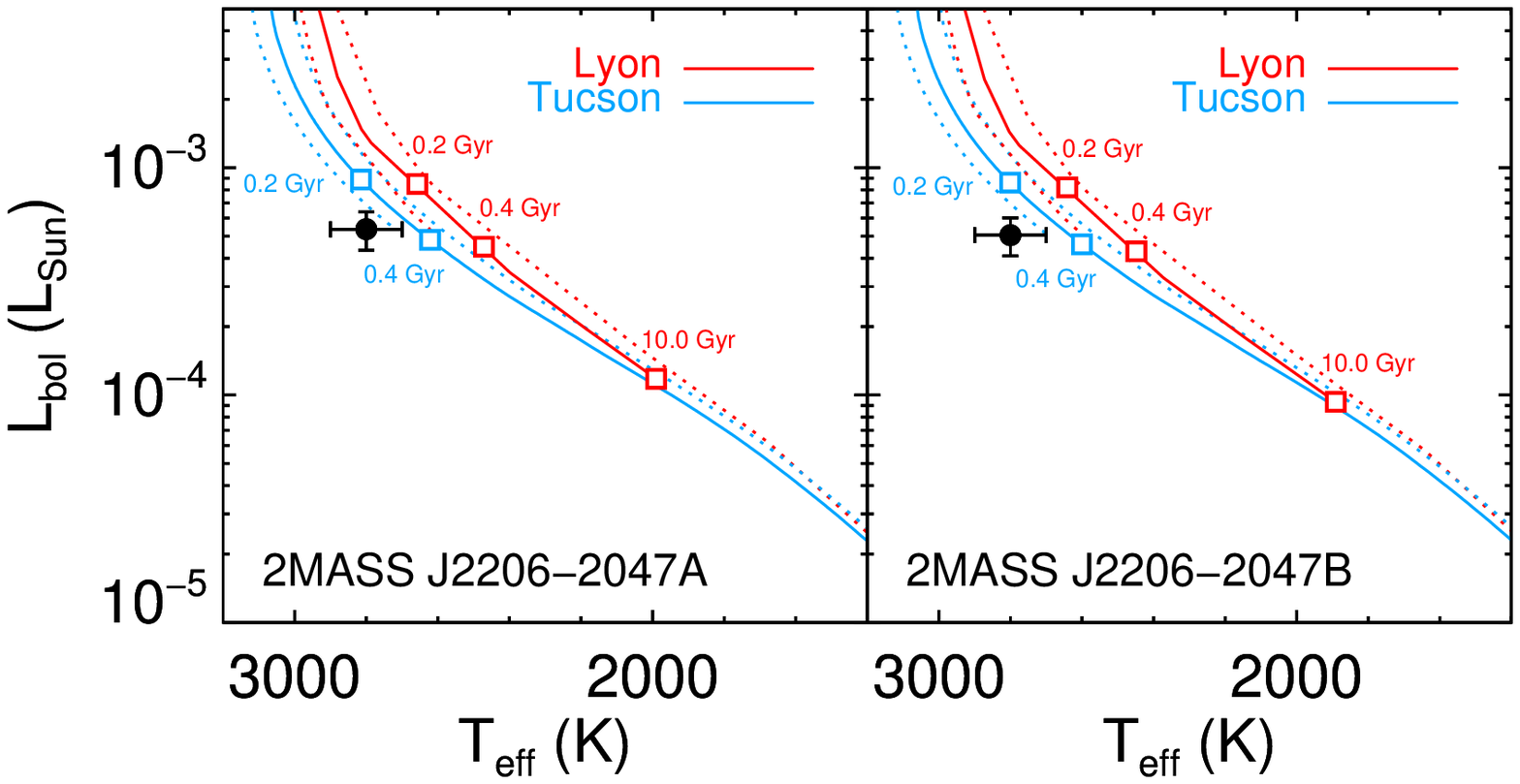}
\caption{\normalsize Hertzsprung-Russell diagram showing isomass lines
  from evolutionary models for the mass of \nameA\ (left) and \nameB\
  (right) with dotted lines encompassing the 1$\sigma$ mass
  uncertainties. The open squares demarcate the median and 1$\sigma$
  confidence limits on the evolutionary model-derived age of \name.
  Our derived effective temperature from spectral synthesis fitting of
  the integrated light spectrum is shown as a filled circle with 100~K
  error bars. The atmospheric model temperature is warmer than both
  sets of evolutionary tracks but is only significantly discrepant
  with the Lyon model tracks. \label{fig:h-r}}
\end{figure*}

\subsection{Temperatures and Surface Gravities \label{sec:teff-logg}}

Without radius measurements for \nameA\ and \nameB, we cannot directly
determine their effective temperatures or surface
gravities.\footnote{Since at least one component of \name\ is
  chromospherically active, it may be possible to estimate one or both
  radii using the technique employed by \citet{2009ApJ...695..310B}
  who measured the rotation period of 2MASSW~J0746425$+$200032A from
  its chromospheric radio emission and combined this with its
  $v\sin(i)$ and orbital inclination ($i$) to derive its radius. This
  method assumes that the orbital and rotation axes are aligned. With
  a $v\sin(i)$ of 22~\kms\ \citep{2002AJ....124..519R}, the active
  component(s) of \name\ is expected to have a rotation period of
  about 280 minutes.} In Section~\ref{sec:atm-fit} we derived these
properties by fitting atmospheric model spectra to the
integrated-light spectrum, and we have also used evolutionary models
to estimate these properties in the same fashion as our model-derived
age and individual masses. The Tucson models give effective
temperatures for \nameA\ and \nameB\ of 2660$^{+90}_{-100}$~K and
2640$^{+90}_{-100}$~K and surface gravities of 5.26$^{+0.09}_{-0.17}$
and 5.26$^{+0.10}_{-0.17}$ (cgs). The Lyon models give systematically
lower but formally consistent temperatures of 2550$^{+90}_{-100}$~K
and 2530$^{+90}_{-100}$~K and surface gravities of
5.17$^{+0.12}_{-0.18}$ and 5.17$^{+0.12}_{-0.18}$ (cgs).  (Note that
the upper confidence limits are likely affected by the same truncation
within our Monte Carlo method as discussed in Section~\ref{sec:qratio}
for the individual masses.)  The differences between the two sets of
models are due to the fact that the Tucson models predict radii that
are 9\% smaller than predicted by Lyon models (Table~\ref{tbl:model}).

Compared to the effective temperature of 2800$\pm$100~K derived from
spectral synthesis fitting, the Tucson models are consistent (at
1.0--1.2$\sigma$), but the Lyon model temperatures are
1.9--2.0$\sigma$ lower.  This is illustrated in Figure~\ref{fig:h-r},
which shows the atmospheric model-derived temperatures in comparison
to the evolutionary tracks on the Hertzsprung-Russell (H-R) diagram.
As a result of this temperature discrepancy, the Tucson and Lyon
model-predicted radii are larger than derived from the atmospheric
model scaling factors by 14\% and 24\%, respectively.  These could be
brought into better agreement if the system were older than the median
model-derived age of 0.4~Gyr, as evolutionary models would predict
smaller radii and thus higher effective temperatures (corresponding to
the 2$\sigma$ lower/upper limits in Table~\ref{tbl:model} for
radii/temperatures).  Finally, atmospheric model fitting did not yield
consistent surface gravity estimates: the dust-free PHOENIX-Gaia
models gave \logg~=~4.5, and the Ames-Dusty models gave \logg~=~6.0
(the maximum allowed by the model grid).  These are respectively lower
and higher than the evolutionary model-derived surface gravities of
\logg~=~5.0--5.4.

\begin{figure*} 
\plotone{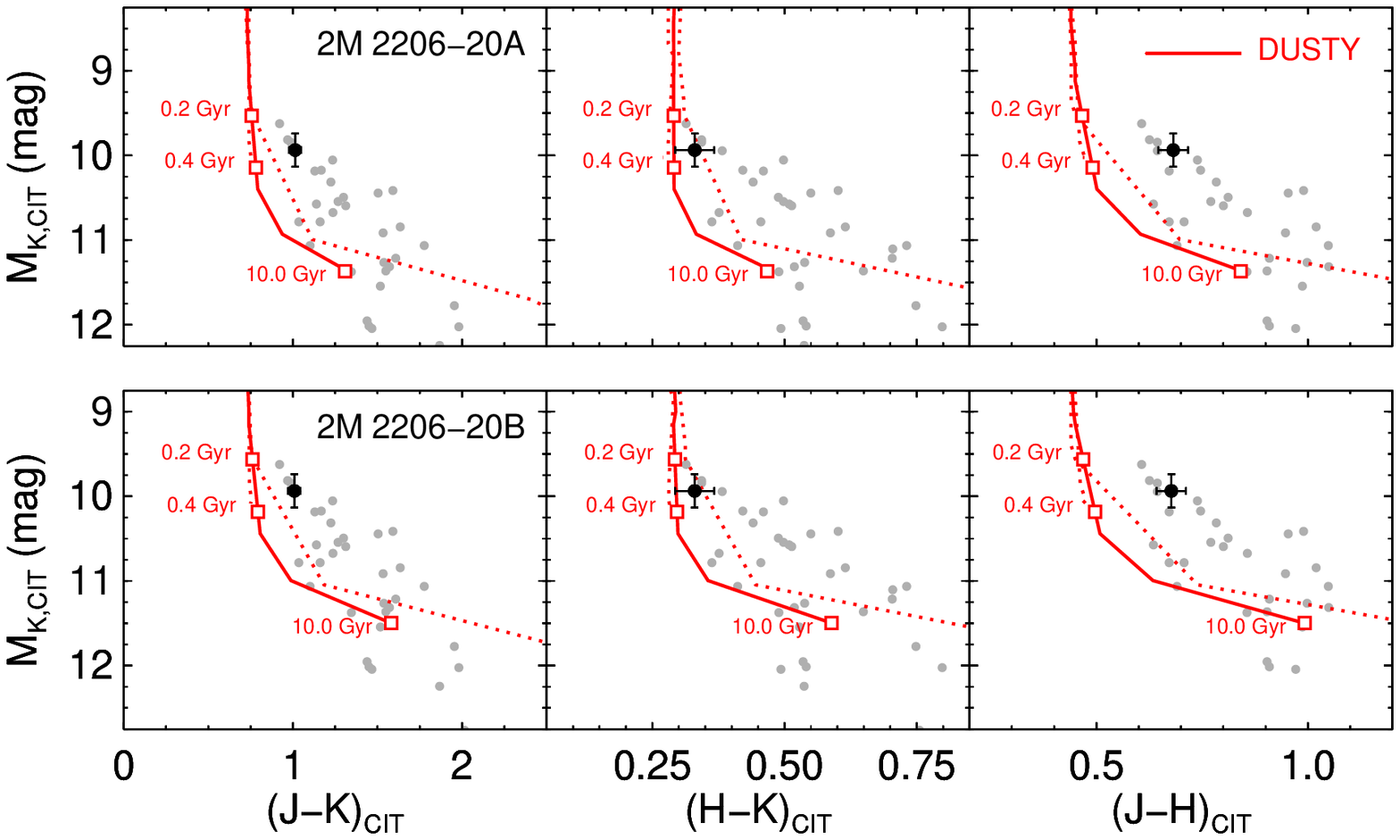}
\caption{ \normalsize Color-magnitude diagrams showing the measured
  photometry of \nameA\ (top) and \nameB\ (bottom) compared to Lyon
  evolutionary tracks ($JHK$ photometry on the CIT system).  The solid
  lines are isomass tracks from the Dusty \citep{2000ApJ...542..464C}
  models with dotted lines encompassing the 1$\sigma$ mass
  uncertainties.  The open squares demarcate the median and 1$\sigma$
  confidence limits on the evolutionary model-derived age of \name.
  Field dwarfs with parallax measurements more precise than 10\% and
  $JHK$ colors more precise than 0.10~mag are shown as filled gray
  circles.  Both components of \name\ have identical colors and are
  typical of field dwarfs.  However, evolutionary models do not
  reproduce the observed colors, with the exception of
  $H-K$. \label{fig:jhk-cmd}}
\end{figure*}

\subsubsection{Comparison to Field Dwarfs \label{sec:teff-field}}

The model-derived effective temperatures for both components of \name\
can be compared to those which have been determined for other objects
of similar spectral type. Temperatures have been estimated in a number
of ways, always relying to some degree on models due to the lack of
direct radius measurements, and we summarize such estimates for late-M
dwarfs in order from most to least model dependent.

\begin{itemize}

\item \emph{Spectral synthesis.} Fitting atmospheric models over a
  very narrow spectral range (2.297--2.310~\micron; $R$~=~42000),
  \citet{2005MNRAS.358..105J} found effective temperatures of 2900~K
  for the two M7--M9 dwarfs in their study. Using a broader spectral
  range (0.7--2.5~\micron; $R$~=~600--3000),
  \citet{2001ApJ...548..908L} found much cooler temperatures of
  2100--2300~K for the five M7--M9 dwarfs in their study. Comparing
  our atmospheric model derived temperature of 2800$\pm$100~K to these
  determinations, it is consistent with the former but $>$~400~K
  inconsistent with the latter. We investigated this discrepancy by
  using a similar spectral fitting approach as
  \citet{2001ApJ...548..908L}, which excludes the 1.5--1.7~\micron\
  portion of the spectrum, weights the spectral regions
  0.7--1.4~\micron\ and 2.0--2.5~\micron\ by a factor of 5 higher than
  the rest of the spectrum, and finally selects the best fitting
  spectrum by eye. When we employed this procedure, we found an
  effective temperature of 2500~K for \nameint, which is consistent
  with the \citet{2001ApJ...548..908L} measurements of M7--M9 dwarfs
  (assuming 100~K uncertainties in both determinations).

\item \emph{Model radii.} For objects with \Lbol\ measurements, the
  nearly flat mass--radius relationship predicted by theoretical
  models for very low-mass stars and brown dwarfs may be utilized to
  estimate \Teff. Adopting an age range of 0.1--10~Gyr,
  \citet{2001ApJ...548..908L} derived effective temperatures of
  1850--2650~K for the five M7--M9 dwarfs in their sample with \Lbol\
  measurements, and this broad range is consistent with our
  model-derived effective temperatures.

\item \emph{Mass benchmarks.} Objects in binaries with measured
  individual luminosities \emph{and} a dynamical total mass enable
  more precise model-derived temperatures and gravities. This is the
  method we have used to determine the effective temperatures of both
  components of \name, and \citet{me-2397a} have previously used an
  identical method to determine the temperature and surface gravity of
  \lhsA\ (M8.0$\pm$0.5): the Tucson models gave 2580$\pm$30~K and
  5.381$^{+0.009}_{-0.014}$ (cgs), and the Lyon models gave
  2470$\pm$30~K and 5.307$^{+0.007}_{-0.008}$ (cgs).  These
  temperatures are consistent with those derived for \name\ from
  evolutionary models, though \lhsA\ is predicted to be 80~K cooler
  due to its slightly lower \Lbol.  Although an indirect comparison,
  we note that our atmospheric model temperature of 2800$\pm$100~K is
  about 200--300~K higher than the evolutionary model derived
  temperature of \lhsA.  The evolutionary model-derived surface
  gravities of \name\ are lower than \lhsA\ and formally inconsistent;
  however, this is easily explained by the higher mass of \lhsA\
  (e.g., Tucson models give 0.0839$^{+0.0007}_{-0.0015}$~\Msun).

\item \emph{Infrared flux method.} The infrared flux method is a
  nearly model-independent way of estimating effective temperature
  that relies on a monochromatic flux measurement in the
  Rayleigh-Jeans tail of an SED as well as a bolometric flux
  measurement \citep{1977MNRAS.180..177B}.
  \citet{2007ApJ...667..527G} used their 24~\micron\ \Spitzer/MIPS
  photometry to determine effective temperatures for nine M7.5--M8.5
  dwarfs by this method, finding temperatures of
  2400--2730~K.\footnote{This range excludes the M8+L7 binary \lhsbin\
    as the companion flux likely contaminates the MIPS measurement
    \citep[see discussion in Section~4.3.1 of][]{me-2397a}.} These
  effective temperatures are in excellent agreement with the
  evolutionary model-derived temperatures for both components of
  \name\ but somewhat lower (70--400~K) than the temperature from
  spectral synthesis fitting.

\end{itemize}

\subsection{Near-Infrared Colors \label{sec:cmd}}

The Lyon evolutionary models provide predictions of the fluxes of
ultracool dwarfs in standard filter bandpasses as a function of model
mass and age.\footnote{The models give $JHK$ photometry on the CIT
  system, and we converted our photometry to this system using the
  relations of \citet{2001AJ....121.2851C}.} We derived the
model-predicted near-infrared colors of both components of \name\ in
the same fashion as the individual masses, effective temperatures, and
age (i.e., using the combined observational constraints of the total
mass and individual luminosities). Figure~\ref{fig:jhk-cmd} shows the
observed colors of \name\ on color--magnitude diagrams in comparison
to model tracks and other field dwarfs. Compared to the observed
colors of the components of \name, only $H-K$ is consistent with the
Dusty models, and both $J-K$ and $J-H$ are about 0.2--0.3~mag redder
than predicted by Dusty. Since field M8 dwarfs have very similar
colors to the components of \name, the Lyon Dusty models will
generally provide inaccurate estimates of the fundamental properties
of late-M dwarfs from their near-infrared colors.


\section{Discussion \label{sec:discuss}}

\subsection{Additional Constraints on the Mass  \label{sec:add-mass}}

We have directly measured the total mass of \name\ to be
0.15$^{+0.05}_{-0.03}$~\Msun. However, the entire range of formally
allowed masses is not consistent with some of its other properties.
For example, at the 1$\sigma$ upper limit in \Mtot, both components of
\name\ would be 0.10~\Msun\ stars, but they would then be 0.7~mag
fainter at $K$ band than the faintest object of comparable mass
\citep[Gl~234B: 0.1034$\pm$0.0035~\Msun,
$M_K$~=~9.26$\pm$0.04;][]{2000A&A...364..217D, 2000A&A...364..665S}.
In contrast, the objects closest in $K$-band brightness to \name\ are
GJ~1245C \citep[0.074$\pm$0.013~\Msun,
$M_K$~=~9.99$\pm$0.04;][]{1999ApJ...512..864H} and \lhsA\
\citep[0.0848$\pm$0.0011~\Msun,\footnote{Note that the mass of \lhsA\
  is derived from a total dynamical mass and evolutionary models.}
$M_K$~=~10.06$\pm$0.07;][]{me-2397a}. Thus, it is more likely that the
components of \name\ have individual masses in this range; in which
case, the total mass would be well below the formally allowed value of
0.20~\Msun.

The 1$\sigma$ lower limit of \Mtot\ = 0.12~\Msun\ corresponds to a
pair of 0.06~\Msun\ brown dwarfs, and masses at or below this value
are also disfavored. \citet{2002AJ....124..519R} found an upper limit
of 0.05~\AA\ for lithium absorption at 6807~\AA\ in the
integrated-light spectrum of \nameint, indicating that both components
have depleted their initial lithium. As discussed by
\citet{1996ApJ...459L..91C}, lithium can only be depleted in objects
more massive than $\approx$0.06~\Msun. Moreover, even more massive
objects require a finite amount of time to become lithium depleted:
according to \citet{1996ApJ...459L..91C}, a 0.070~\Msun\ object takes
0.2~Gyr to destroy 99\% of its initial lithium, with lower mass
objects taking longer. Given the constraint of the individual
luminosities of \name, the low-mass tail of the \Mtot\ distribution
corresponds to young ages. At the median mass of 0.15~\Msun\ the
model-derived age is 0.4~Gyr, and at the 1$\sigma$ lower bound of
0.12~\Msun\ the age is 0.2~Gyr. Below this 1$\sigma$ limit, the
components of \name\ would be inconsistent with the lithium
non-detection: (1) they would have had insufficient time to destroy
their initial lithium, and/or (2) they should be low enough mass that
they would never destroy any lithium. Thus, regardless of the precise
location of the lithium-fusing boundary, the formally allowed low-mass
tail of the \Mtot\ distribution is not physically plausible.

\subsection{Direct Measurement of the Mass Ratio  \label{sec:qdirect}}

Since the components of \name\ are nearly identical, testing models
using our measured total mass is straightforward. However, future
measurements may constrain the binary's mass ratio directly. Given
that the flux ratio is so near unity, the mass ratio would not be
feasible to measure from astrometric monitoring of the photocenter
since the center-of-light would be imperceptibly different from the
center-of-mass. Thus, the mass ratio must be determined through radial
velocity monitoring. This is also challenging as the binary is
currently approaching $\Delta{V}$ = 0~\kms\ and will not reach the
next peak in the radial velocity curve for $P/4 \approx$~9~years, in
2018. Until then, the radial velocities of the two components will
remain below 1.4~\kms. Since the velocity of each component must be
measured to 7\% in order to determine the mass ratio to 10\%, radial
velocity measurements with a precision better than 0.1~\kms\ are
needed. This is at the limit of state-of-the-art techniques using
current instrumentation \citep{2007ApJ...666.1198B} but are well
within reach of future near-infrared spectrographs with precision
goals of 1~m~s$^{-1}$ \citep{2008SPIE.7014E..31J}.


\section{Conclusions}

We have determined the orbit of the M8+M8 binary \name\ using relative
astrometry spanning 8.3~years of its 35$^{+6}_{-5}$~year orbit. The
astrometry and corresponding errors used to derive this orbit were
thoroughly examined through Monte Carlo simulations, using PSF
reference sources for the AO images. The resulting best-fit orbit has
a reduced $\chi^2$ of 1.07 and total mass of
0.15$^{+0.05}_{-0.03}$~\Msun. Because the orbit only contributes 2.0\%
to the mass error, the uncertainty in the dynamical mass is dominated
by the 9.1\% error in the parallax, which translates into an
asymmetric $^{+32}_{-22}$\% mass error. Although this mass is
sufficiently precise to perform interesting model tests, a more
precise parallax would provide even stronger tests and would remove
the large ambiguity in the characterization of the system (i.e.,
whether it is composed of young brown dwarfs or old stars).

We have used evolutionary models to derive the properties of \name\
using Monte Carlo methods developed in previous studies
\citep[e.g.,][]{2008ApJ...689..436L, 2009ApJ...692..729D}. Both the
Tucson and Lyon Dusty models give an age for the system of
0.4$^{+9.6}_{-0.2}$~Gyr. The median age is somewhat young because the
median total mass is somewhat low given the individual luminosities;
however, the 1$\sigma$ upper bound extends to the maximum allowed age
(10~Gyr). This model-derived age is consistent with \nameint's space
motion and its chromospheric activity.

We also derived the near-infrared colors of both components of \name\
from the Lyon models and compared them to our observations. We found
that the model $J-H$ and $J-K$ colors were significantly
(0.2--0.3~mag) bluer than observed, while the model $H-K$ colors were
in good agreement, suggestive of an important opacity source missing
at $J$ band in the Dusty models (or else systematic errors that cancel
out for $H-K$). In any case, our observations show that masses and/or
ages derived from the Dusty evolutionary models on the
color--magnitude diagram will be in error for objects such as \name.

Our effective temperature determinations from evolutionary models are
in very good agreement with \Teff\ determinations for other M7.5--M8.5
dwarfs: (1) from the infrared flux method \citep{2007ApJ...667..527G},
and (2) from a similar mass benchmark system including the M8 dwarf
\lhsA\ \citep{me-2397a}. We also derived effective temperatures for
both components from atmospheric model fitting of the integrated-light
spectrum, which is made possible by their essentially identical fluxes
and colors. We found that these temperatures (2800$\pm$100~K) are
warmer than predicted by evolutionary models and are most discrepant
(2$\sigma$) with the Lyon Dusty models (n.b., the surface boundary
condition for the Lyon evolutionary models is determined by the same
Dusty atmospheric models as we used for spectral synthesis fitting).
This modest discrepancy may be caused by systematic errors in the
atmospheric models, which use a maximal limiting case in the treatment
of dust and incomplete line lists. Alternatively, the discrepancy
could be explained if the system were somewhat older than the median
age of 0.4~Gyr inferred from its luminosity and mass, as this would
cause evolutionary model radii to be smaller and the derived effective
temperatures warmer. In such a scenario, the true mass would be in the
high-mass tail of the \Mtot\ distribution, corresponding to a larger
distance to the system, which can be tested directly with an improved
parallax measurement.

Stars at the bottom of the main sequence experience much of the same
atmospheric physics as the warmest brown dwarfs and extrasolar planets
because of the presence of dust in their photospheres. The
characterization of such objects largely relies on theoretical models
that must accurately describe the behavior of this dust as well as the
opacity due to millions of molecular transitions
\citep[e.g.,][]{2006MNRAS.368.1087B}. Dynamical mass measurements for
ultracool binaries like \name\ provide the critical benchmarks for
testing and improving these models. The future holds many more such
benchmarks as ongoing orbital monitoring efforts have only begun to
yield new dynamical masses from the large samples of ultracool dwarfs
discovered by wide field surveys nearly a decade ago.


\acknowledgments

We gratefully acknowledge the Keck AO team for their exceptional
efforts in bringing the AO system to fruition. It is a pleasure to
thank Antonin Bouchez, David LeMignant, Marcos van Dam, Al Conrad,
Randy Campbell, Carolyn Parker, Joel Aycock, Hien Tran, and the Keck
Observatory staff for assistance with the observations. We are very
thankful for the contribution of Peter Tuthill in establishing
aperture masking at Keck. We are grateful to Brian Cameron for making
available his NIRC2 distortion solution, C\'{e}line Reyl\'{e} for
customized Besan\c{c}on Galaxy models, and Adam Burrows and Isabelle
Baraffe for providing finely gridded evolutionary models. We have
benefited from discussions with Michael Cushing about theoretical
models and Thierry Forveille about space motions and orbit fitting
using \orbit. We are indebted to Katelyn Allers for assistance in
obtaining IRTF/SpeX data.
Our research has employed the 2MASS data products; NASA's
Astrophysical Data System; the SIMBAD database operated at CDS,
Strasbourg, France; the SpeX Prism Spectral Libraries, maintained by
Adam Burgasser at \texttt{http://www.browndwarfs.org/spexprism}; and
the M, L, and T~dwarf compendium housed at
\texttt{http://DwarfArchives.org} and maintained by Chris Gelino, Davy
Kirkpatrick, and Adam Burgasser \citep{2003IAUS..211..189K,
  2004AAS...205.1113G}.
TJD and MCL acknowledge support for this work from NSF grant
AST-0507833, and MCL acknowledges support from an Alfred P. Sloan
Research Fellowship.
Finally, the authors wish to recognize and acknowledge the very
significant cultural role and reverence that the summit of Mauna Kea has
always had within the indigenous Hawaiian community.  We are most
fortunate to have the opportunity to conduct observations from this
mountain.

{\it Facilities:} \facility{Keck II Telescope (LGS AO, NIRC2)},
\facility{\HST\ (WFPC2)}, \facility{VLT (NACO)},
\facility{Gemini-North Telescope (Hokupa`a), \facility{IRTF (SpeX)}}



\begin{thebibliography}{57}
\expandafter\ifx\csname natexlab\endcsname\relax\def\natexlab#1{#1}\fi

\bibitem[{Allard {et~al.}(2001)Allard, {Hauschildt}, {Alexander}, {Tamanai}, \&
  {Schweitzer}}]{2001ApJ...556..357A}
Allard, F., {Hauschildt}, P.~H., {Alexander}, D.~R., {Tamanai}, A., \&
  {Schweitzer}, A. 2001, \apj, 556, 357

\bibitem[{Barber {et~al.}(2006)Barber, {Tennyson}, {Harris}, \&
  {Tolchenov}}]{2006MNRAS.368.1087B}
Barber, R.~J., {Tennyson}, J., {Harris}, G.~J., \& {Tolchenov}, R.~N. 2006,
  \mnras, 368, 1087

\bibitem[{Berger {et~al.}(2009)Berger, {Rutledge}, {Phan-Bao}, {Basri},
  {Giampapa}, {Gizis}, {Liebert}, {Mart{\'{\i}}n}, \&
  {Fleming}}]{2009ApJ...695..310B}
Berger, E., {et~al.} 2009, \apj, 695, 310

\bibitem[{Blackwell \& {Shallis}(1977)}]{1977MNRAS.180..177B}
Blackwell, D.~E., \& {Shallis}, M.~J. 1977, \mnras, 180, 177

\bibitem[{Blake {et~al.}(2007)Blake, {Charbonneau}, {White}, {Marley}, \&
  {Saumon}}]{2007ApJ...666.1198B}
Blake, C.~H., {Charbonneau}, D., {White}, R.~J., {Marley}, M.~S., \& {Saumon},
  D. 2007, \apj, 666, 1198

\bibitem[{Bouy {et~al.}(2003)Bouy, {Brandner}, {Mart{\'{\i}}n}, {Delfosse},
  {Allard}, \& {Basri}}]{2003AJ....126.1526B}
Bouy, H., {Brandner}, W., {Mart{\'{\i}}n}, E.~L., {Delfosse}, X., {Allard}, F.,
  \& {Basri}, G. 2003, \aj, 126, 1526

\bibitem[{Brott \& {Hauschildt}(2005)}]{2005ESASP.576..565B}
Brott, I., \& {Hauschildt}, P.~H. 2005, in ESA Special Publication, Vol. 576,
  The Three-Dimensional Universe with Gaia, ed. C.~{Turon}, K.~S. {O'Flaherty},
  \& M.~A.~C. {Perryman}, 565

\bibitem[{Burrows {et~al.}(1997)Burrows, {Marley}, {Hubbard}, {Lunine},
  {Guillot}, {Saumon}, {Freedman}, {Sudarsky}, \&
  {Sharp}}]{1997ApJ...491..856B}
Burrows, A., {et~al.} 1997, \apj, 491, 856

\bibitem[{Carpenter(2001)}]{2001AJ....121.2851C}
Carpenter, J.~M. 2001, \aj, 121, 2851

\bibitem[{Chabrier {et~al.}(2000)Chabrier, {Baraffe}, {Allard}, \&
  {Hauschildt}}]{2000ApJ...542..464C}
Chabrier, G., {Baraffe}, I., {Allard}, F., \& {Hauschildt}, P. 2000, \apj, 542,
  464

\bibitem[{Chabrier {et~al.}(1996)Chabrier, {Baraffe}, \&
  {Plez}}]{1996ApJ...459L..91C}
Chabrier, G., {Baraffe}, I., \& {Plez}, B. 1996, \apjl, 459, L91

\bibitem[{Close {et~al.}(2002)Close, {Siegler}, {Potter}, {Brandner}, \&
  {Liebert}}]{2002ApJ...567L..53C}
Close, L.~M., {Siegler}, N., {Potter}, D., {Brandner}, W., \& {Liebert}, J.
  2002, \apjl, 567, L53

\bibitem[{Costa {et~al.}(2006)Costa, {M{\'e}ndez}, {Jao}, {Henry},
  {Subasavage}, \& {Ianna}}]{2006AJ....132.1234C}
Costa, E., {M{\'e}ndez}, R.~A., {Jao}, W.-C., {Henry}, T.~J., {Subasavage},
  J.~P., \& {Ianna}, P.~A. 2006, \aj, 132, 1234

\bibitem[{Crifo {et~al.}(2005)Crifo, {Phan-Bao}, {Delfosse}, {Forveille},
  {Guibert}, {Mart{\'{\i}}n}, \& {Reyl{\'e}}}]{2005A&A...441..653C}
Crifo, F., {Phan-Bao}, N., {Delfosse}, X., {Forveille}, T., {Guibert}, J.,
  {Mart{\'{\i}}n}, E.~L., \& {Reyl{\'e}}, C. 2005, \aap, 441, 653

\bibitem[{Cushing {et~al.}(2008)Cushing, {Marley}, {Saumon}, {Kelly}, {Vacca},
  {Rayner}, {Freedman}, {Lodders}, \& {Roellig}}]{2008ApJ...678.1372C}
Cushing, M.~C., {et~al.} 2008, \apj, 678, 1372

\bibitem[{Cushing {et~al.}(2005)Cushing, {Rayner}, \&
  {Vacca}}]{2005ApJ...623.1115C}
Cushing, M.~C., {Rayner}, J.~T., \& {Vacca}, W.~D. 2005, \apj, 623, 1115

\bibitem[{Cushing {et~al.}(2004)Cushing, {Vacca}, \&
  {Rayner}}]{2004PASP..116..362C}
Cushing, M.~C., {Vacca}, W.~D., \& {Rayner}, J.~T. 2004, \pasp, 116, 362

\bibitem[{Cutri {et~al.}(2003)Cutri, {Skrutskie}, {van Dyk}, {Beichman},
  {Carpenter}, {Chester}, {Cambresy}, {Evans}, {Fowler}, {Gizis}, {Howard},
  {Huchra}, {Jarrett}, {Kopan}, {Kirkpatrick}, {Light}, {Marsh}, {McCallon},
  {Schneider}, {Stiening}, {Sykes}, {Weinberg}, {Wheaton}, {Wheelock}, \&
  {Zacarias}}]{2mass}
Cutri, R.~M., {et~al.} 2003, {2MASS All Sky Catalog of point sources.} (The
  IRSA 2MASS All-Sky Point Source Catalog, NASA/IPAC Infrared Science
  Archive.~http://irsa.ipac.caltech.edu/applications/Gator/)

\bibitem[{Dahn {et~al.}(2002)}]{2002AJ....124.1170D}
Dahn, C.~C., {et~al.} 2002, \aj, 124, 1170

\bibitem[{Delfosse {et~al.}(2000)Delfosse, {Forveille}, {S{\'e}gransan},
  {Beuzit}, {Udry}, {Perrier}, \& {Mayor}}]{2000A&A...364..217D}
Delfosse, X., {Forveille}, T., {S{\'e}gransan}, D., {Beuzit}, J.-L., {Udry},
  S., {Perrier}, C., \& {Mayor}, M. 2000, \aap, 364, 217

\bibitem[{Dupuy {et~al.}(2009{\natexlab{a}})Dupuy, {Liu}, \&
  {Ireland}}]{2009ApJ...692..729D}
Dupuy, T.~J., {Liu}, M.~C., \& {Ireland}, M.~J. 2009{\natexlab{a}}, \apj, 692,
  729

\bibitem[{Dupuy {et~al.}(2009{\natexlab{b}})Dupuy, {Liu}, \&
  {Ireland}}]{me-2397a}
---. 2009{\natexlab{b}}, \apj, 699, 168

\bibitem[{Faherty {et~al.}(2009)Faherty, {Burgasser}, {Cruz}, {Shara},
  {Walter}, \& {Gelino}}]{2009AJ....137....1F}
Faherty, J.~K., {Burgasser}, A.~J., {Cruz}, K.~L., {Shara}, M.~M., {Walter},
  F.~M., \& {Gelino}, C.~R. 2009, \aj, 137, 1

\bibitem[{Forveille {et~al.}(1999)Forveille, {Beuzit}, {Delfosse}, {Segransan},
  {Beck}, {Mayor}, {Perrier}, {Tokovinin}, \& {Udry}}]{1999A&A...351..619F}
Forveille, T., {et~al.} 1999, \aap, 351, 619

\bibitem[{Gautier {et~al.}(2007)Gautier, {Rieke}, {Stansberry}, {Bryden},
  {Stapelfeldt}, {Werner}, {Beichman}, {Chen}, {Su}, {Trilling}, {Patten}, \&
  {Roellig}}]{2007ApJ...667..527G}
Gautier, III, T.~N., {et~al.} 2007, \apj, 667, 527

\bibitem[{Gelino {et~al.}(2004)Gelino, {Kirkpatrick}, \&
  {Burgasser}}]{2004AAS...205.1113G}
Gelino, C.~R., {Kirkpatrick}, J.~D., \& {Burgasser}, A.~J. 2004, \baas, 205

\bibitem[{Ghez {et~al.}(2008)Ghez, {Salim}, {Weinberg}, {Lu}, {Do}, {Dunn},
  {Matthews}, {Morris}, {Yelda}, {Becklin}, {Kremenek}, {Milosavljevic}, \&
  {Naiman}}]{2008ApJ...689.1044G}
Ghez, A.~M., {et~al.} 2008, \apj, 689, 1044

\bibitem[{Gizis {et~al.}(2000)Gizis, {Monet}, {Reid}, {Kirkpatrick}, {Liebert},
  \& {Williams}}]{2000AJ....120.1085G}
Gizis, J.~E., {Monet}, D.~G., {Reid}, I.~N., {Kirkpatrick}, J.~D., {Liebert},
  J., \& {Williams}, R.~J. 2000, \aj, 120, 1085

\bibitem[{Golimowski {et~al.}(2004)}]{gol04}
Golimowski, D.~A., {et~al.} 2004, \aj, 127, 3516

\bibitem[{Guenther \& {Wuchterl}(2003)}]{2003A&A...401..677G}
Guenther, E.~W., \& {Wuchterl}, G. 2003, \aap, 401, 677

\bibitem[{Henry {et~al.}(1999)Henry, {Franz}, {Wasserman}, {Benedict},
  {Shelus}, {Ianna}, {Kirkpatrick}, \& {McCarthy}}]{1999ApJ...512..864H}
Henry, T.~J., {Franz}, O.~G., {Wasserman}, L.~H., {Benedict}, G.~F., {Shelus},
  P.~J., {Ianna}, P.~A., {Kirkpatrick}, J.~D., \& {McCarthy}, Jr., D.~W. 1999,
  \apj, 512, 864

\bibitem[{Henry \& {McCarthy}(1993)}]{1993AJ....106..773H}
Henry, T.~J., \& {McCarthy}, Jr., D.~W. 1993, \aj, 106, 773

\bibitem[{Ibukiyama \& {Arimoto}(2002)}]{2002A&A...394..927I}
Ibukiyama, A., \& {Arimoto}, N. 2002, \aap, 394, 927

\bibitem[{Jones {et~al.}(2005)Jones, {Pavlenko}, {Viti}, {Barber}, {Yakovina},
  {Pinfield}, \& {Tennyson}}]{2005MNRAS.358..105J}
Jones, H.~R.~A., {Pavlenko}, Y., {Viti}, S., {Barber}, R.~J., {Yakovina},
  L.~A., {Pinfield}, D., \& {Tennyson}, J. 2005, \mnras, 358, 105

\bibitem[{Jones {et~al.}(2008)Jones, {Rayner}, {Ramsey}, {Henry}, {Dent},
  {Montgomery}, {Vick}, {Ives}, {Egan}, {Lunney}, {Rees}, {Webster}, {Tinney},
  \& {Liu}}]{2008SPIE.7014E..31J}
Jones, H.~R.~A., {et~al.} 2008, in Society of Photo-Optical Instrumentation
  Engineers (SPIE) Conference Series, Vol. 7014, Society of Photo-Optical
  Instrumentation Engineers (SPIE) Conference Series

\bibitem[{Kirkpatrick(2003)}]{2003IAUS..211..189K}
Kirkpatrick, J.~D. 2003, in Proceedings of IAU Symposium 211: Brown Dwarfs, ed.
  E.~Martin, 189

\bibitem[{Kirkpatrick {et~al.}(1991)Kirkpatrick, {Henry}, \&
  {McCarthy}}]{1991ApJS...77..417K}
Kirkpatrick, J.~D., {Henry}, T.~J., \& {McCarthy}, Jr., D.~W. 1991, \apjs, 77,
  417

\bibitem[{Krist(1995)}]{1995ASPC...77..349K}
Krist, J. 1995, in Astronomical Society of the Pacific Conference Series,
  Vol.~77, Astronomical Data Analysis Software and Systems IV, ed. R.~A.
  {Shaw}, H.~E. {Payne}, \& J.~J.~E. {Hayes}, 349

\bibitem[{Leggett {et~al.}(2001)Leggett, {Allard}, {Geballe}, {Hauschildt}, \&
  {Schweitzer}}]{2001ApJ...548..908L}
Leggett, S.~K., {Allard}, F., {Geballe}, T.~R., {Hauschildt}, P.~H., \&
  {Schweitzer}, A. 2001, \apj, 548, 908

\bibitem[{Lenzen {et~al.}(2003)Lenzen, {Hartung}, {Brandner}, {Finger},
  {Hubin}, {Lacombe}, {Lagrange}, {Lehnert}, {Moorwood}, \&
  {Mouillet}}]{2003SPIE.4841..944L}
Lenzen, R., {et~al.} 2003, in Presented at the Society of Photo-Optical
  Instrumentation Engineers (SPIE) Conference, Vol. 4841, Society of
  Photo-Optical Instrumentation Engineers (SPIE) Conference Series, ed.
  M.~{Iye} \& A.~F.~M. {Moorwood}, 944--952

\bibitem[{Liu {et~al.}(2008)Liu, {Dupuy}, \& {Ireland}}]{2008ApJ...689..436L}
Liu, M.~C., {Dupuy}, T.~J., \& {Ireland}, M.~J. 2008, \apj, 689, 436

\bibitem[{Monet {et~al.}(1992)Monet, {Dahn}, {Vrba}, {Harris}, {Pier},
  {Luginbuhl}, \& {Ables}}]{1992AJ....103..638M}
Monet, D.~G., {Dahn}, C.~C., {Vrba}, F.~J., {Harris}, H.~C., {Pier}, J.~R.,
  {Luginbuhl}, C.~B., \& {Ables}, H.~D. 1992, \aj, 103, 638

\bibitem[{Rayner {et~al.}(2003)Rayner, {Toomey}, {Onaka}, {Denault},
  {Stahlberger}, {Vacca}, {Cushing}, \& {Wang}}]{2003PASP..115..362R}
Rayner, J.~T., {Toomey}, D.~W., {Onaka}, P.~M., {Denault}, A.~J.,
  {Stahlberger}, W.~E., {Vacca}, W.~D., {Cushing}, M.~C., \& {Wang}, S. 2003,
  \pasp, 115, 362

\bibitem[{Reid {et~al.}(2002)Reid, Kirkpatrick, {Liebert}, {Gizis}, {Dahn}, \&
  {Monet}}]{2002AJ....124..519R}
Reid, I.~N., Kirkpatrick, J.~D., {Liebert}, J., {Gizis}, J.~E., {Dahn}, C.~C.,
  \& {Monet}, D.~G. 2002, \aj, 124, 519

\bibitem[{Robin {et~al.}(2003)Robin, {Reyl{\'e}}, {Derri{\`e}re}, \&
  {Picaud}}]{2003A&A...409..523R}
Robin, A.~C., {Reyl{\'e}}, C., {Derri{\`e}re}, S., \& {Picaud}, S. 2003, \aap,
  409, 523

\bibitem[{Rousset {et~al.}(2003)Rousset, {Lacombe}, {Puget}, {Hubin},
  {Gendron}, {Fusco}, {Arsenault}, {Charton}, {Feautrier}, {Gigan}, {Kern},
  {Lagrange}, {Madec}, {Mouillet}, {Rabaud}, {Rabou}, {Stadler}, \&
  {Zins}}]{2003SPIE.4839..140R}
Rousset, G., {et~al.} 2003, in Presented at the Society of Photo-Optical
  Instrumentation Engineers (SPIE) Conference, Vol. 4839, Society of
  Photo-Optical Instrumentation Engineers (SPIE) Conference Series, ed. P.~L.
  {Wizinowich} \& D.~{Bonaccini}, 140--149

\bibitem[{S{\'e}gransan {et~al.}(2000)S{\'e}gransan, {Delfosse}, {Forveille},
  {Beuzit}, {Udry}, {Perrier}, \& {Mayor}}]{2000A&A...364..665S}
S{\'e}gransan, D., {Delfosse}, X., {Forveille}, T., {Beuzit}, J.-L., {Udry},
  S., {Perrier}, C., \& {Mayor}, M. 2000, \aap, 364, 665

\bibitem[{Seifahrt {et~al.}(2008)Seifahrt, {R{\"o}ll}, {Neuh{\"a}user},
  {Reiners}, {Kerber}, {K{\"a}ufl}, {Siebenmorgen}, \&
  {Smette}}]{2008A&A...484..429S}
Seifahrt, A., {R{\"o}ll}, T., {Neuh{\"a}user}, R., {Reiners}, A., {Kerber}, F.,
  {K{\"a}ufl}, H.~U., {Siebenmorgen}, R., \& {Smette}, A. 2008, \aap, 484, 429

\bibitem[{Simons \& {Tokunaga}(2002)}]{mkofilters1}
Simons, D.~A., \& {Tokunaga}, A. 2002, \pasp, 114, 169

\bibitem[{Tegmark {et~al.}(2004)}]{2004PhRvD..69j3501T}
Tegmark, M., {et~al.} 2004, \prd, 69, 103501

\bibitem[{Tokunaga {et~al.}(2002)Tokunaga, {Simons}, \& {Vacca}}]{mkofilters2}
Tokunaga, A.~T., {Simons}, D.~A., \& {Vacca}, W.~D. 2002, \pasp, 114, 180

\bibitem[{Vacca {et~al.}(2003)Vacca, {Cushing}, \&
  {Rayner}}]{2003PASP..115..389V}
Vacca, W.~D., {Cushing}, M.~C., \& {Rayner}, J.~T. 2003, \pasp, 115, 389

\bibitem[{van Dam {et~al.}(2006)}]{2006PASP..118..310V}
van Dam, M.~A., {et~al.} 2006, \pasp, 118, 310

\bibitem[{West {et~al.}(2008)West, {Hawley}, {Bochanski}, {Covey}, {Reid},
  {Dhital}, {Hilton}, \& {Masuda}}]{2008AJ....135..785W}
West, A.~A., {Hawley}, S.~L., {Bochanski}, J.~J., {Covey}, K.~R., {Reid},
  I.~N., {Dhital}, S., {Hilton}, E.~J., \& {Masuda}, M. 2008, \aj, 135, 785

\bibitem[{Wizinowich {et~al.}(2006)}]{2006PASP..118..297W}
Wizinowich, P.~L., {et~al.} 2006, \pasp, 118, 297

\bibitem[{Wood \& {Oswalt}(1998)}]{1998ApJ...497..870W}
Wood, M.~A., \& {Oswalt}, T.~D. 1998, \apj, 497, 870

\bibitem[{Zapatero~Osorio {et~al.}(2007)Zapatero~Osorio, {Mart{\'{\i}}n},
  {B{\'e}jar}, {Bouy}, {Deshpande}, \& {Wainscoat}}]{2007ApJ...666.1205Z}
Zapatero~Osorio, M.~R., {Mart{\'{\i}}n}, E.~L., {B{\'e}jar}, V.~J.~S., {Bouy},
  H., {Deshpande}, R., \& {Wainscoat}, R.~J. 2007, \apj, 666, 1205

\end{thebibliography}
\end{document}